\documentclass[12pt,a4paper]{article}
\usepackage{jheppub}
\usepackage{dsfont}
\usepackage{amssymb,amsmath}
\usepackage{epsfig}
\usepackage{epstopdf}
\usepackage{latexsym}
\usepackage{graphicx}
\usepackage{subfig}
\usepackage{booktabs}
\usepackage{bbm}
\pdfoutput=1
\makeatletter
\def\@fpheader{\relax}
\makeatother

\newcommand{\be}{\begin{equation}}
\newcommand{\ee}{\end{equation}}

\def\E{{\cal E}}
\def\F{{\cal F}}
\def\cM{{\cal M}}
\def\cN{{\cal N}}

\def\S{{\cal S}}
\def\A{{\cal A}}

\def\L{{\cal L}}
\def\O{{\cal O}}
\def\J{{\cal J}}
\def\S{{\cal S}}
\def\G{{\cal G}}
\def\cH{{\cal H}}
\def\T{{\cal T}}
\def\E{{\cal E}}
\def\W{{\cal W}}
\def\N{{\cal N}}
\def\P{{\cal P}}
\def\Z{{\cal Z}}
\def\B{{\cal B}}

\newcommand\bigDiamond{\mathop{\mathpalette\bigDi@mond\relax}}

\title{Emergent Horizons and Causal Structures in Holography}
\author{Avik Banerjee$^a$, Arnab Kundu$^a$, Sandipan Kundu$^b$}
\affiliation{$^a$Theory Division, Saha Institute of Nuclear Physics, 1/AF Bidhannagar, Kolkata 700064, India.}
\affiliation{$^b$Department of Physics, Cornell University, Ithaca, New York, 14853, USA.}
\emailAdd{avik.banerjee [at] saha.ac.in} 
\emailAdd{arnab.kundu [at] saha.ac.in} 
\emailAdd{kundu [at] cornell.edu}

\abstract{The open string metric arises kinematically in studying fluctuations of open string degrees of freedom on a D-brane. An observer, living on a probe D-brane, can send signals through the spacetime by using such fluctuations on the probe, that propagate in accordance with a metric which is conformal to the open string metric. Event horizons can emerge in the open string metric when one considers a D-brane with an electric field on its worldvolume. Here, we emphasize the role of and investigate, in details, the causal structure of the resulting open string event horizon and demonstrate, among other things, its close similarities to an usual black hole event horizon in asymptotically AdS-spaces.  To that end, we analyze relevant geodesics, Penrose diagrams and various  causal holographic observables for a given open string metric. For analytical control, most of our calculations are performed in an asymptotically AdS$_3$-background, however, we argue that the physics is qualitatively the same in higher dimensions. We also discuss how this open string metric arises from an underlying D-brane configuration in string theory.}

\begin{document}

\maketitle
\flushbottom

\section{Introduction \& Discussions}

The basic mathematical statement of gauge-string duality is an equivalence of the path integral (or the partition function, in Euclidean signature) of the quantum field theory with the path integral (or partition function) of supergravity, which is usually an Einstein-gravity theory with various matter fields, typically having its origin in string theory. Schematically, without worrying about a rigorous definition, the equivalence reads:
\begin{eqnarray}
\Z_{\rm sugra} = \Z_{\rm QFT} \ , \label{gsdual1}
\end{eqnarray}
where $\Z_{\rm QFT}$ usually corresponds to a large $N_c$ gauge theory, with only adjoint degrees of freedom and $\Z_{\rm sugra}$ corresponds to a particular supergravity theory that can be obtained as the low energy limit of closed string theory.  The statement in (\ref{gsdual1}) is best understood {\it on-shell}, {\it e.g.}~when the LHS is evaluated on a solution (and thus a saddle point of the path integral) of supergravity. The fluctuation modes thereof will couple to the metric corresponding to the saddle point and will yield the following path integral:
\begin{eqnarray}
&& \Z_{\rm sugra + fluc} = \Z_{\rm sugra} \ \Z_{\rm fluc} \ , \\
&& \Z_{\rm fluc} = {\rm exp} \left( \int \frac{i}{2} d\phi \wedge \ast d \phi  - \frac{i}{2} F \wedge \ast F - \frac{i}{2} H \wedge \ast H + {\rm interactions} \right) \ ,
\end{eqnarray}
where $\phi$, $F$ and $H$ correspond to the scalar, vector and tensor perturbations, respectively. Additionally, there will be interaction terms that we need not specify here. Clearly, the Hodge star operation above is defined with respect to a metric, denoted by $G$ henceforth, that solves supergravity and defines a causal structure. Given this, in the context of gauge-string duality, one can analyze the manifestation of the bulk causal structure on the boundary theory correlator (observables in general). This has a long history, and has been explored in details in {\it e.g.}~\cite{Fidkowski:2003nf, Hubeny:2005qu, Hubeny:2006yu, Hubeny:2012ry}.

So far the discussion is without any ``quark"-like matter field. To add to it a matter sector that transforms in the fundamental representation of the gauge group, one needs to introduce the so-called ``open string degrees of freedom"\cite{Karch:2002sh}. The first step is to consider a ``probe limit" in which the number of fundamental flavour, denoted by $N_f$, satisfies $N_f \ll N_c$, and hence does not gravitationally back-react on the geometry. The dynamics of the probe D-branes embedded appropriately in a given $10$-dimensional supergravity background is governed by the Dirac-Born-Infeld (DBI) action. The statement in (\ref{gsdual1}) generalizes to:
\begin{eqnarray}
\Z_{\rm sugra + DBI} = \Z_{\rm Adj + Fund} \ , \label{sugdbi_adjflav}
\end{eqnarray}
where the symbols are self-explanatory. The idea here is, given a $G$ that already solves supergravity, to find a ``saddle point" configuration of the DBI-theory. Now, once this saddle point is obtained, one can analyze the fluctuation modes around it. Subsequently, similar to the discussion above, there will be a symmetric rank $2$ tensor that defines the kinetic terms of the corresponding fluctuation modes (in this case, those can be scalar, vector and spinor)\cite{Kundu:2013eba}. This symmetric two tensor, denoted by $\S$ henceforth and will be explained better in the subsequent section, therefore defines a causal structure obeyed by the fluctuation modes on the probe brane. In this article we analyze properties of geodesics (spacelike, null and timelike) in the geometry characterized by $\S$. For an earlier discussion on the emergence of an effective causal structure, see {\it e.g.}~\cite{Gibbons:2001gy}. Note that, as discussed in \cite{Gibbons:2001gy}, the emergent causal structure from a Dirac-Born-Infeld theory does not allow the corresponding fluctuations travel faster than the background gravitons.

In this article we consider a particular configuration on the probe brane such that the corresponding open string metric (osm) develops an event horizon even when the original (super)gravity background is horizon-less. This is simply obtained by exciting an electric field on the probe worldvolume, which then sets the horizon of the corresponding osm geometry. Now we begin briefly summarizing our results. The {\it emergent} horizon structure of the osm is subtle if one wants to view it as a solution of Einstein-gravity sourced by a particular matter field. This would have to violate one of the energy conditions ({\it e.g.}~Weak Energy Condition or Null Energy Condition depending on the dimension)\footnote{In practice, given the open string geometry, we will check whether the corresponding Ricci tensor obeys such conditions. Ordinarily, the Ricci tensor is related to the energy-momentum tensor, and hence a condition on the Ricci tensor translates, {\it via} Einstein's equation, into an energy condition.} and thus we can safely state that the resulting metric and therefore the causal structure is {\it in principle} different from that obtained from Einstein-gravity. An equivalent statement would be: the osm horizon is a {\it kinematical} property which emerges from a particular configuration, rather than a dynamical one that extremises an action. However, what we find here is in close qualitative similarity to usual AdS-BH geometries.

We analyze properties of geodesics, specially spacelike and null geodesics, to study the corresponding causal structure. We reveal, among other things, numerous similarities to the standard black hole geometries in an AdS-background, {\it e.g.}~a spacelike geodesic anchored at two points on the conformal boundary, in an osm-geometry, reaches arbitrarily close to the horizon which is an unstable orbit itself. There are, however, technical differences of subtle nature: the Penrose diagram for a BTZ geometry and the corresponding Penrose diagram of an AdS$_3$-osm geometry are identical, but the structure of the singularity is different. In particular, for AdS$_3$-osm geometry, two null rays emanating from the two sides of the Kruskally-extended patch at $T=0$, $T$ being the Kruskal time, would fall into the singularity before meeting each other. Furthermore, based on the causal structure, we explore various observables that are defined in the bulk osm-geometry from a purely causal point of view, and also observe similarities to an AdS-BH geometry. Most of our explicit calculations are performed in AdS$_3$-osm geometry that allows a lot of analytical control. We also provide a top-down, D-brane model that would yield this osm that we have studied extensively.

The crucial difference is in the identification of the osm event-horizon area with a physical quantity, despite the close resemblance of a purely thermal physics as observed in {\it e.g.}~\cite{Mateos:2006nu, Albash:2006ew}. It is clear, based on the discussion on energy conditions, the area of the osm event horizon would not necessarily have a monotonically increasing property, and in fact examples exist where they have explicit non-monotonic behaviour\cite{WIP}. On the other hand, given the proposal for thermodynamic free energy in \cite{Alam:2012fw} and further explored in \cite{Banerjee:2015cvy}, it is straightforward to obtain a covariant formula for thermal entropy. This would, however, involve extrinsic curvatures and can readily be perceived to take complicated mathematical form and thus we leave this for future work.

This article is divided in the following sections: First we introduce the open string metric in section 2 and discuss the context in which they appear, in details. Subsequently, in section 3 we study the causal structure in details, by analyzing geodesics in the given osm-background. Section 4 is devoted to a brief discussion on the various energy conditions in the context of an osm, then we discuss causal observables in section 5. In the next section, we discuss a stringy embedding of the AdS$_3$-osm and finally, we have provided various details of our calculations in five appendices.

\section{The open string metric}

It is well-known that the osm arises when one considers fluctuation modes on the worldvolume of a probe brane that is embedded in a background geometry.\footnote{Note that here we work in the context of Gauge-gravity duality, but the notion of an open string metric is more general. See {\it e.g.}~\cite{Seiberg:1999vs} where an open string propagating in a non-commutative geometry gets naturally equipped with the open string metric.} This geometry is typically a solution of Type IIA/IIB or $11$-dimensional supergravity that arises in the low energy limit of closed string theory or M-theory. This background is dual to a large-$N$ (super)Yang-Mills, or a Chern-Simons matter theory. The probe brane corresponds to introducing a fundamental (flavour) matter sector in this large-$N$ gauge theory.

Let us begin by elaborating on the emergence of the open string metric in more details. In order to do that, we will consider the simplest example: D$3$-D$7$ system. However, the main conclusion does not depend on the details of the system and hence it is true in general. We start with $N_c$ coincident D$3$ branes in type IIB supergravity with the action schematically written as:
\be
S_{\rm sugra} \equiv N_c^2\ {\cal{S}}_{\rm sugra} \left[\phi,G \right]\ , \label{sugra}
\ee
where $G$ is the metric and $\phi$ is the collection of all other supergravity fields. Note that we have written the factor of $N_c^2$ explicitly in action (\ref{sugra}) for later convenience. The supergravity partition function now can be written as:
\be
\Z_{\rm sugra} = \int D\left[\phi \right]D \left [G\right] e^{-N_c^2 {\cal{S}}_{\rm sugra}[\phi,G]}\ .
\ee
In the limit $N_c\rightarrow \infty$, we can perform a saddle point approximation and the above partition function can be replaced by a sum over semiclassical minima of the supergravity action $S_{\rm sugra}$, with appropriate boundary conditions. In the example we are discussing here, there is a unique vacuum supergravity solution carrying D$3$ brane charge, namely, AdS$_5\times$S$^5$ and hence in the leading order in $N_c^2$, we obtain:
\be
\Z_{\rm sugra} = e^{-N_c^2\ {\cal{S}}_{\rm sugra}^{(0)}}\ , \label{approx}
\ee
where ${\cal{S}}_{\rm sugra}^{(0)}$ is the on-shell classical supergravity action. There are two types of corrections to the partition function (\ref{approx}): (i) string theory correction --- in other words, the so-called $\alpha'$ corrections --- to the supergravity action, and (ii) corrections from quantum loops computed around the classical supergravity solution. We restrict to the case of $l_s/R\rightarrow 0$, where $l_s \sim \sqrt{\alpha'}$ is the string length and $R$ is the AdS radius, for which we can safely ignore the stringy corrections. We are more interested in the quantum correction to the above saddle point approximation which is responsible for $1/N_c^2$ correction. This will schematically yield:
\be
\Z_{\rm sugra} = \int D[\delta \phi]D[\delta G] e^{-N_c^2\left( {\cal{S}}_{\rm sugra}^{(0)} + {\cal{S}}_{\rm sugra}^{(2)}[\delta \phi, \delta G] +...\right)}\ , \label{approx1}
\ee
where ${\cal{S}}_{\rm sugra}^{(2)}[\delta \phi, \delta G]$ is the quadratic action of the perturbations around the supergravity solution.\footnote{There is no linear term in (\ref{approx1}) because we are expanding around the classical solution.} Perturbations $\delta \phi$ and  $\delta G$ see the background metric $G$ and hence the causal structure of the metric $G$ determines how fast information can be sent through the bulk.

Let us now introduce $N_f$ space-filling D$7$ branes embedded in the AdS$_5\times$S$^5$ background. The dynamics of the D$7$ branes is determined by the Dirac-Born-Infeld (DBI) action (supplemented by the Wess-Zumino term, when necessary) of the form:
\begin{eqnarray} \label{DBI}
S_{\rm DBI} = -N_f \tau_7 \int d \xi^8  e^{-\Phi} \sqrt{- {\rm det} \left( {\varphi}^\star\left[G + B \right] + \left(2 \pi \alpha' \right) F \right)}  \ ,
\end{eqnarray}
where $\{\Phi, G, B\}$ are the supergravity data in the string-frame consisting of the dilaton, the metric and the NS-NS $2$-form, respectively. The RR-forms generally appear in the Wess-Zumino part which we are not explicitly writing. For the case we are considering $\phi\equiv \{\Phi, B\}=0$ for the classical supergravity solution. On the other hand, $d\xi^8$ denotes the integration over the probe worldvolume coordinates, $\varphi^\star$ denotes the pull-back and $F$ denotes the $U(1)$-gauge field that lives on the worldvolume of the probe. The $N_f$ D$7$ branes wrap the full AdS$_5$ along with a S$^3$ $\subset$ S$^5$.  For convenience, we will set $2\pi \alpha' = 1$. Let us now write the action (\ref{DBI}) schematically in the following form:
\be
S_{\rm DBI} = g_s N_f N_c \ \S_{\rm DBI} \left[G,\phi; F, \theta_a\right] \ ,
\ee
where, $g_s$ is the string coupling that will eventually set the 't Hooft coupling, $\theta_a$ with $a=1,2$ represents the profile of the D$7$ branes. So the full partition function now can be written as:
\be
\Z_{\rm sugra+DBI} = \int D[\phi] D[G] D[F] D[\theta_a]e^{-N_c^2 {\cal{S}}_{\rm sugra}[\phi,G] - g_s N_f N_c \S_{\rm DBI} [G,\phi; F, \theta_a]} \ .
\ee
In the limit, $N_c\rightarrow \infty$, $N_c N_f\rightarrow \infty$ and $N_f/N_c\rightarrow 0$, we can again perform a saddle point approximation
\be
\Z_{\rm sugra+DBI} = e^{-N_c^2{\cal{S}}_{\rm sugra}^{(0)} - g_s N_f N_c \S_{\rm DBI}^{(0)}}\ , \label{sgdbi}
\ee 
where, $\S_{\rm DBI}^{(0)}$ is the on-shell DBI action obtained by solving the classical equations of motion for $\theta_a$. For simplicity, we will assume that 
\be
\theta_a^{(0)} = \text{constant}\ , \qquad a=1,2
\ee
minimizes the DBI action.\footnote{For our present purpose, this is only a notational convenience, we are not loosing any physics due to this.} Let us now consider fluctuations around the classical solution. The full partition function can be approximated in the following way:
\be
\Z_{\rm sugra+DBI} = \Z_{\rm classical} \ \Z_{\rm fluc}\ ,
\ee
$\Z_{\rm classical}$ is the classical part of the partition function (leading saddle point approximation):
\be
\Z_{\rm classical} = e^{-N_c^2{\cal{S}}_{\rm sugra}^{(0)} - g_s N_f N_c \S_{\rm DBI}^{(0)} - g_s N_f^2 \S_{\rm back-reac}^{(1)} + \O \left (N_f/N_c \right)} \ .
\ee
The leading classical term comes from the supergravity action. The subleading term is the contribution of the DBI action. The sub-subleading term $N_f^2 \S_{\rm back-reac}^{(1)}$ comes from the back-reaction of the D$7$ branes on the geometry. Note that the back-reaction of the DBI action on the geometry in general can be ignored in the probe limit $N_f/N_c\ll 1$, however, we have kept it in the above equation just to remind ourselves that the leading back-reaction term can be large compare to the contributions of the fluctuations.

In the quadratic order the fluctuation part $\Z_{\rm fluc}$ can be decomposed into the supergravity part and DBI part:
\be
\Z_{\rm fluc}=\int D[\delta \phi]D[\delta G]e^{-N_c^2{\cal{S}}_{\rm sugra}^{(2)}[\delta \phi, \delta G]+...} \int D[\delta F] D[\delta \theta_a]e^{- g_s N_c N_f{\cal{S}}_{DBI}^{(2)}[\delta F, \delta \theta_a]+..} \ ,\label{partition}
\ee
where ${\cal{S}}_{\rm sugra}^{(2)}[\delta \phi, \delta G]$  and $g_s N_c N_f {\cal{S}}_{\rm DBI}^{(2)}[\delta F, \delta \theta_a]$ are the quadratic actions of the perturbations around the classical solution.\footnote{Note that in equation (\ref{partition}), we do not need to include the corrections to the supergravity background because of backreaction of the D$7$ branes since they are further suppressed by factors of $N_f/N_c$.} The dots in equation (\ref{partition}) represent contributions that are suppressed by factors of $N_f/N_c$. Also note that, in principle the DBI action contains fluctuations of $\delta \phi$ and $\delta G$, however these contributions come with additional factors $N_f/N_c$ and hence we will ignore them.

Let us now look at the fluctuations of the DBI action in equation (\ref{partition}). It can be decomposed into a scalar part and a vector part in the following way. The scalar fluctuations are the transverse fluctuations of the D$7$ brane embedding:
\begin{eqnarray}
\theta_a = \theta_a^{(0)} + \varphi_a  \ .
\end{eqnarray}
The vector fluctuations correspond to the fluctuations of the classical gauge field on the probe
\begin{eqnarray}
F_{ab} = F^{(0)}_{ab} + \F_{ab} \ ,
\end{eqnarray}
where $F^{(0)}$ is the classical gauge field on the D$7$ branes. Now one can study the dynamics of fluctuations. In doing so, it becomes immediately clear that the kinetic term of the corresponding fluctuation mode, irrespective of the type of fluctuation\cite{Kundu:2013eba, Kundu:2015qda, Banerjee:2015cvy}, takes the following schematic form:
\begin{eqnarray}
 \S_{\rm scalar} & = & - \frac{\kappa}{2} \int d\xi^8\left(\frac{{\rm det} G}{{\rm det} \S} \right)^{1/4} \sqrt{-{\rm det} \S} \,  \S^{ab} \, \partial_a \varphi^i \,  \partial_b \varphi^i  + \ldots  , \label{fscalar} \\
 \S_{\rm vector} & = & - \frac{\kappa}{4}  \int d\xi^8 \left(\frac{{\rm det} G}{{\rm det} \S} \right)^{1/4} \sqrt{-{\rm det} \S} \, \S^{ab}\S^{cd} \, \F_{ac} \F_{bd} + \ldots \ . \label{fvector} 
\end{eqnarray}
In (\ref{fscalar})-(\ref{fvector}), $\kappa$ denotes an overall constant, the details of which is not relevant for us. The fields $\varphi^i$, $\F$ represent the various fluctuation modes and the indices $a,b$ represent the worldvolume coordinates on the probe.\footnote{It is also possible to have spinor fluctuations come from a supersymmetric counter-part of the DBI action, which schematically consists of a standard Volkov-Akulov type term
\begin{eqnarray}
S_{\rm VA} =  - N_f \tau_p \int d^{8} \xi  \left( - {\rm det} \left[{\varphi}^\star\left[G  \right] + \left(2 \pi \alpha' \right) F + i \bar{\psi} \gamma \nabla \psi \right]\right)^{1/2} , \label{dbi2}
\end{eqnarray}
where the $\gamma$ matrices satisfy anti-commutation relation with respect to $ {\varphi}^\star\left[G  \right]$: $\left\{\gamma_a, \gamma_b\right\} = 2{\varphi}^\star\left[G  \right]_{ab}$.
}
In (\ref{fscalar})-(\ref{fvector}), $\S$ is the open string metric, defined as
\begin{eqnarray}
\S = \varphi^\star G - \left(F \left(\varphi^\star G\right)^{-1} F \right)  \ . \label{osmdef}
\end{eqnarray}
We have shown only the kinetic parts of the fluctuation Lagrangian; since other potential terms will not affect our discussion for now. Before moving further, let us note the following: The Lagrangian density corresponding to (\ref{fscalar}) and (\ref{fvector}) can be written in a more canonical form:
\begin{eqnarray}
&& \sqrt{- {\rm det} \tilde{\S}} \ \tilde{\S}^{ab} \left(\partial_a \varphi \right)  \left(\partial_b \varphi \right) \ , \quad {\rm and} \quad  \sqrt{- {\rm det} \tilde{\S}} \ \tilde{\S}^{ab} \tilde{\S}^{cd} \F_{ac} \F_{bd} \ , \label{canokin} \\
&& {\rm where} \quad \tilde{\S}_{\rm nc} = \Omega  \S_{\rm nc} \ , \label{confosm}
\end{eqnarray}
and $\Omega$ needs to be determined for each case, separately. Here $\tilde{\S}_{\rm nc}$ and $\S_{\rm nc}$ correspond to the non-compact parts of the metric components. Note that, the conformal factor in {\it e.g.}~(\ref{confosm}) will not affect the effective temperature that is obtained from the osm itself.

Let us now go back to the partition function (\ref{partition}). The equation (\ref{partition}) now we can be rewritten as:
\be
\Z_{\rm fluc} = \int D[\delta \phi]D[\delta G]e^{-N_c^2{\cal{S}}_{\rm sugra}^{(2)}[\delta \phi, \delta G]+...} \int D[\delta \F]D[\delta \varphi_a] e^{- g_s N_c N_f( \S_{\rm scalar}[\varphi_a]+ \S_{\rm vector}[\F])+..}\ .\label{partition1}
\ee
By looking at the quadratic action, we can conclude the following: An observer can send a signal through the bulk in two separate ways: (i) he/she can perturb the supergravity fields ($\delta G, \delta \phi$) to send a signal and propagation of this signal will be controlled by the bulk metric $G$, (ii) he/she can also use the D$7$ brane fluctuations $\varphi$ or $\F_{ab}$ to send a signal, however this signal propagates through the spacetime in accordance with a metric which is conformal to the open string metric.\footnote{It's conformal to the open string metric but not exactly the open string metric because of the presence of the non-trivial factor in (\ref{fscalar})-(\ref{fvector}). } Therefore, both the bulk metric $G$ and the open string metric $\S$ control the propagation of information through the spacetime and hence the study of the causal structure of the open string metric is of importance. A reasonable guess is that one would not be able to send signal faster by using the D$7$ brane fluctuations $\varphi$ or $\F_{ab}$ because that will violate causality, and this is indeed the case, as argued in {\it e.g.}~\cite{Gibbons:2001gy}. We will also confirm this by performing some explicit computations. It is also well-known by now\cite{Kundu:2013eba, Kundu:2015qda, Kim:2011qh, Sonner:2012if} that one induces an event horizon in this geometry by having a classically non-vanishing electric field on the probe worldvolume, that subsequently induces a Schwinger pair production\cite{Sonner:2013mba}. This suggests that there are regions where one can send signal by using the supergravity fields but not by using fluctuations $\varphi$ or $\F_{ab}$.

Before we proceed, let us also note that from partition function (\ref{partition1}), one can obtain different correlation functions of the perturbations. For example, from the supergravity part of the action we get $\langle \delta G \delta G\rangle \sim 1/N_c^2$; similarly from the DBI part of the action leads to $\langle \varphi_a \varphi_a \rangle \sim 1/N_fN_c$. Higher order terms in the action will contribute to quantum loop corrections which are subleading in the limits $N_c\gg 1$, $N_c N_f\gg 1$.

So far we have not said anything about the AdS/CFT correspondence. The AdS/CFT correspondence teaches us that our supergravity set up is dual to a $(3+1)$-dimensional ${\cal N}=4$ supersymmetric SU($N_c$) Yang-Mills theory with an additional $N_f$ number of  ${\cal N} = 2$ supersymmetric hypermultiplets transforming in the fundamental representation of the gauge group. This gauge theory can be well approximated by the supergravity set up we described before in the limit $N_c\gg 1$, $N_c N_f \gg 1$, $N_f/N_c \ll 1$ and $\lambda=g_{YM}^2 N_c \gg 1$. In particular, we have
\begin{eqnarray}
\Z_{\rm sugra + DBI} = \Z_{\rm Adj + Fund} \ ,
\end{eqnarray}
which is what we have introduced in (\ref{sugdbi_adjflav}).

One final remark: In this article we are concerned about the causal structure of the open string metric for a specific configuration, in which an event horizon emerges to what the open string degrees of freedom couple, it is not present in the supergravity background. We will frequently consider the open string metric itself, and ignore the conformal factor since the causal structure is not sensitive to the conformal factor.

\section{Probing the geometry: Causal structure}

In gauge-gravity duality, the classical relativistic probes that correspond to various non-local operators in the dual field theory, are extremal surfaces of various co-dimensions. Examples include space-like geodesics. These yield the two-point correlation function of operators with large dimensions\cite{Banks:1998dd}. More precisely, in the WKB approximaiton:
\begin{eqnarray}
\langle \O(x) \O(y)  \rangle = {\rm exp} \left( - m \L_{\rm geo}[x, y] \right) \ ,
\end{eqnarray}
where $m$ is the mass of bulk field, $\L_{\rm geo}[x,y]$ is the (renormalized) geodesic length and $\O$ corresponds to the dual field theory operator.

There can certainly be extremal surfaces of other co-dimensions, {\it e.g.}~Wilson-Maldacena loops that are determined by extremal area string world sheets\cite{Maldacena:1998im}, or entanglement entropy of a region $\A$ in the boundary that is determined by a co-dimension two extremal area surface ending on $\partial \A$\cite{Ryu:2006bv}. Thus, the properties of extremal surfaces play a crucial role in determining how the dual CFT-data is encoded in the geometry, see {\it e.g.}~\cite{Hubeny:2012ry} for an extensive study of extremal surfaces in asymptotically locally AdS-spaces.

In this article, the geometries that we consider, are indeed asymptotically AdS; however, they do not arise as solutions to Einstein-gravity. Thus, while the standard AdS/CFT-dictionary of identifying extremal geodesics with dual field theory quantities is unclear in our case, it is certainly a well-posed question purely from a gravitational point of view and hence, by virtue of the duality, should correspond to a well-defined observable in the dual field theory. Thus, in this section, we will analyze features of geodesics in an open string metric geometry, with the hope that it corresponds to physical properties for the fundamental matter sector in the dual field theory, even though we will not make the correspondence any more precise at this point.

\subsection{Open string geometry of AdS$_3$}

We start with the AdS$_{2+1}$ metric in the following form:
\begin{equation}
ds^2= \frac{1}{u^2} \left( - dt^2 + dx^2 + du^2 \right) \ , \label{ads3def}
\end{equation}
where AdS radius has been set to unity. The conformal boundary is located at $u \to 0$ and the infrared of the geometry is located at $u \to \infty$. In a BTZ geometry, the infrared will contain a horizon that sets the temperature of the dual $(1+1)$-dimensional conformal field theory. We will offer some comments regarding the dual field theory in a later section.

Now we will consider introducing ``space-filling" probe branes in the AdS$_3$-background.\footnote{We will make the assumption that such space-filling embedding exists. This, for the most part of our purpose, is a simplifying assumption that does not necessarily loose any physical information.} Specifically to introduce an event horizon in the osm, we will excite the following gauge potential: 
\begin{eqnarray}
A_x = - E t + a_x(u) \quad {\rm with} \quad  F = dA \ . \label{gauge}
\end{eqnarray}
The physics of this fundamental matter sector is rather intuitive: Since we applied an electric field, there will be pair-creation even in the absence of explicit charge density and this will drive a flavour-current. The corresponding current, denoted by $j \sim \left( \partial \L_{\rm DBI} / \partial a_x'\right)$, is essentially given by the first integral of motion for the field $a_x(u)$. See {\it e.g.}~\cite{Karch:2007pd, Albash:2007bq} for more details on a representative example of embedding D$7$-brane in AdS$_5\times S^5$-background.

Now, using the definition in (\ref{osmdef}), for the background in (\ref{ads3def}) and the gauge field in (\ref{gauge}) the corresponding osm is calculated to be:
\begin{eqnarray}
ds_{\rm osm}^2 & = & -\frac{1}{u^2}\left(1-\frac{u^4}{u_*^4}\right)d\tau^2+\left(\frac{1}{u^2}+\frac{1}{u_*^2}\right) dx^2 + \frac{1}{u^2}\left(\frac{1}{1-\frac{u^2}{u_*^2}}\right)du^2 \ , \label{osmetric} \\
d\tau & = &  dt - \frac{E j u^3}{\sqrt{\left( E^2 u^4 -1 \right) \left( j^2 u^2 -1 \right)}} du \ , \label{tautdef}
\end{eqnarray}
with
\begin{equation}
E = \frac{1}{u_*^2} \ , \qquad  j = \frac{1}{u_*} = \sqrt{E} \ .
\end{equation}
With reference to (\ref{confosm}), also note that
\begin{eqnarray}
\tilde{\S} = \Omega \S \ , \quad \Omega = \left( 1 + \frac{u^2}{u_*^2} \right)^{-1} \ . \label{confosm3}
\end{eqnarray}
Clearly, the osm in (\ref{osmetric}) inherits a structure similar to a black hole geometry, with an effective Hawking temperature:
\begin{equation}
T_{\rm eff} =  \frac{E^{1/2}}{\sqrt{2}\pi} \ .
\end{equation}
%

\subsection{A comparative study of geodesics}

We will begin with a comparative analysis of what is known in BTZ-background, which is given by
\begin{eqnarray}
ds^2 = -\frac{1}{u^2}\left(1-\frac{u^2}{u_{\rm H}^2}\right)dt^2 + \frac{1}{u^2} dx^2+\frac{1}{u^2}\left(\frac{1}{1-\frac{u^2}{u_{\rm H}^2}}\right)du^2 \ , \label{btzdef}
\end{eqnarray}
 and what we get in AdS$_3$-osm geometry given in (\ref{osmetric}). Towards that end, let us note the temperature corresponding to (\ref{btzdef}) is: $T=1/\left(2\pi u_{\rm H} \right)$, and let us set $T_{\rm eff} = T$ to relate the parameters $u_{\rm H}$ with $u_*$ as: $u_{\rm H} = u_* / \left(\sqrt{2} \right)$. Thus, in the dual field theory, we are comparing the fundamental matter sector at $T_{\rm eff}$ with purely adjoint sector at $T = T_{\rm eff}$. In the following section, we will review the results discussed in \cite{Fidkowski:2003nf} for BTZ-background and borrow their technique for analyzing the same with the osm-geometry.

\subsubsection{Radial null geodesic}

For completeness and convenience, we have summarized the geodesic equations in appendix \ref{sec:geod}, and we will use the relevant formulae whenever necessary. We begin by discussing properties of null geodesics. Geodesic paths can be found by extremizing the action
\begin{equation}
S = \int ds~ \S_{\mu \nu}\frac{dx^\mu}{ds}\frac{dx^\nu}{ds} \ . 
\end{equation}
where $s$ is an affine parameter, $\S_{\mu\nu}$ are the metric coefficients, and $x^{\mu}$ are the space-time coordinates.

We will focus on $x = {\rm constant}$--slices. A null geodesic on this slice in the BTZ-background is described by
\begin{eqnarray}
\frac{dt}{du}=\frac{1}{1-\frac{u^2}{u_{\rm H}^2}} \quad \implies \quad t(u) = t_0 + \frac{u_{\rm H}}{2}\ln \left(\frac{u_{\rm H} + u}{u_{\rm H} - u}\right) \ ,
\end{eqnarray}
which satisfies the boundary condition: $t(u=0) = t_0$. A null geodesic on a similar slice of the AdS$_3$-osm in (\ref{osmetric}) can be described by
\begin{eqnarray}
&& \frac{d\tau}{du} = \frac{\left(1-\frac{u^2}{u_*^2}\right)^{-1}}{\sqrt{1+\frac{u^2}{u_*^2}}} \quad \implies \quad  \tau(u) = \tau_0 + \frac{u_*}{2\sqrt{2}}\ln \left(\frac{\sqrt{u^2+u_*^2}+u\sqrt{2}}{\sqrt{u^2+u_*^2}-u\sqrt{2}}\right)\ , \\
&& {\rm alternatively}, \quad t(u) = \tau_0 + u_* - \sqrt{u^2+u_*^2} + \frac{u_*}{\sqrt{2}} \ln \left(\frac{\sqrt{2(u^2+u_*^2)}+u+u_*}{(\sqrt{2}+1)(u_*-u)}\right) \ . \nonumber \\ 
\end{eqnarray}
The boundary condition is, evidently, $\tau(u=0) = \tau_0$. In the second line above, we have used the relation in (\ref{tautdef}).

So far what we have discussed, should be interpreted as an analysis ``outside the horizon", following the approach in \cite{Fidkowski:2003nf}. To consider the ``inside" of the black hole, let us introduce the complexified time coordinate:
\begin{equation}
t = t_{\rm L} + i t_{\rm E} \ ,
\end{equation}
where $t_{\rm L}$ and $t_{\rm E}$ denote times on the Lorentzian and Euclidean slices respectively. In the BTZ-background we obtain
\begin{equation}\label{BTZnull}
t(u) = t_0 + \frac{u_{\rm H}}{2} \ln \left(\frac{u + u_{\rm H}}{u - u_{\rm H}}\right) - i \frac{\pi u_{\rm H}}{2} \ .
\end{equation}
The the last term is simply $i\beta /4$, where  $T=1/\beta = 1/(2\pi  u_{\rm H})$ is the Hawking temperature of the BTZ black hole. The equation (\ref{BTZnull}) implies that a null geodesic which starts at the boundary at $t_0 = 0$ reaches the singularity $u =\infty$ at $t_\infty = -i\frac{\pi u_{\rm H}}{2}$, which is purely imaginary. Hence the geodesic hits the singularity at the centre in the Penrose diagram that can subsequently be drawn as a square.

Let us now look at the AdS$_3$-osm. For null geodesics that cross the horizon of AdS$_3$-osm we get
\begin{eqnarray}
&& \tau(u) = \tau_0 + \frac{u_*}{2\sqrt{2}}\ln \left(\frac{\sqrt{u^2+u_*^2}+u\sqrt{2}}{\sqrt{u^2+u_*^2}-u\sqrt{2}}\right) - i \frac{\pi u_*}{2\sqrt{2}} \ , \quad {\rm alternatively}, \\
&&  t(u) = t_0 +  u_*-  \sqrt{u^2+u_*^2} + \frac{u_*}{\sqrt{2}} \ln \left(\frac{\sqrt{2(u^2+u_*^2)} + u + u_*}{(\sqrt{2}+1)(u-u_*)}\right) - i \frac{\pi u_*}{2\sqrt{2}} \ .
\end{eqnarray}
The imaginary part is similarly $i \beta_{\rm eff}/4$, which is consistent with $T_{\rm eff} = 1/( \sqrt{2}\pi u_*)$. From the above equation we get that a null geodesic starting at the boundary at $t_0 = 0$ reaches the singularity $u =\infty$ at $t_\infty =-\infty -i\frac{\pi u_*}{2\sqrt{2}}$, which is not purely imaginary. This is equivalently seen in the $\tau(u)$ coordinate, in which $u \to \infty$ yields $\tau \to c_0 - i\frac{\pi u_*}{2\sqrt{2}}$, where $c_0$ is a finite, but non-zero constant.

\begin{figure}[t]
\centering 
\includegraphics[width=0.8\textwidth]{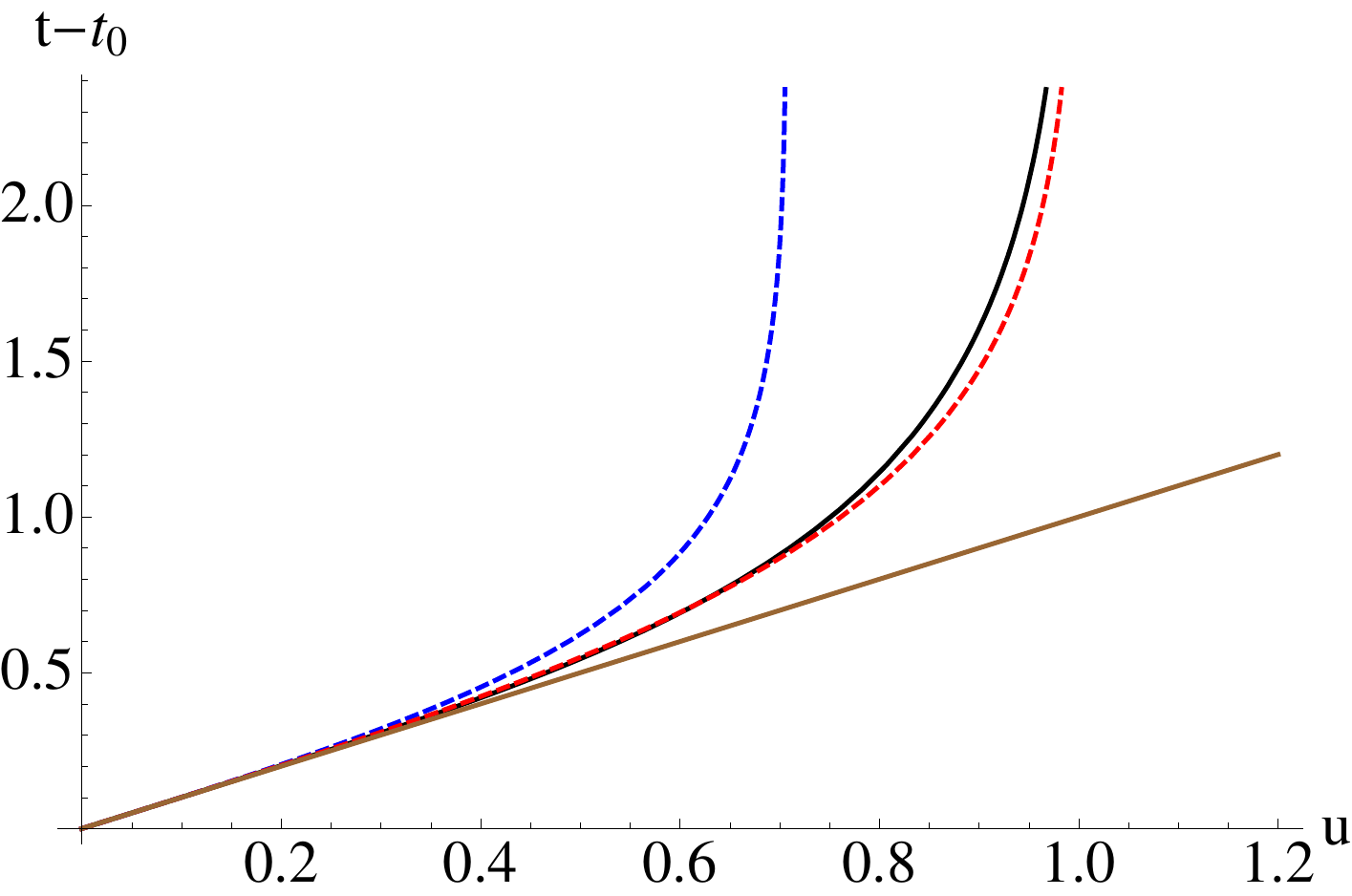} 
\caption{Null geodesics: $t(u)-t_0$ along an ingoing radial null geodesic which starts from the boundary $u =0$ at $t = t_0$ for AdS$_3$-osm (solid black line), BTZ with $u_{\rm H}=u_*/\sqrt{2}$ (dashed blue line), BTZ with  $u_{\rm H}=u_*$ (dashed red line) and AdS$_3$ (solid brown line). We have set $u_*=1$. }
\label{ngout}
\end{figure}

Thus, the global feature of the corresponding Penrose diagram will be different compared to the BTZ-one. Specifically, the Penrose diagram will not be square-shaped, which we will demonstrate later. It is also interesting to note that in order to reach the singularity at ${\rm Re}[t(u\rightarrow \infty)]=0$, the null geodesic has to start at $t_0=\infty$. The results of these discussions are pictorially summarized in figures \ref{ngout} and \ref{ngin}.
\begin{figure}[ht!] \label{ngin}
\centering
\includegraphics[width=0.8\textwidth]{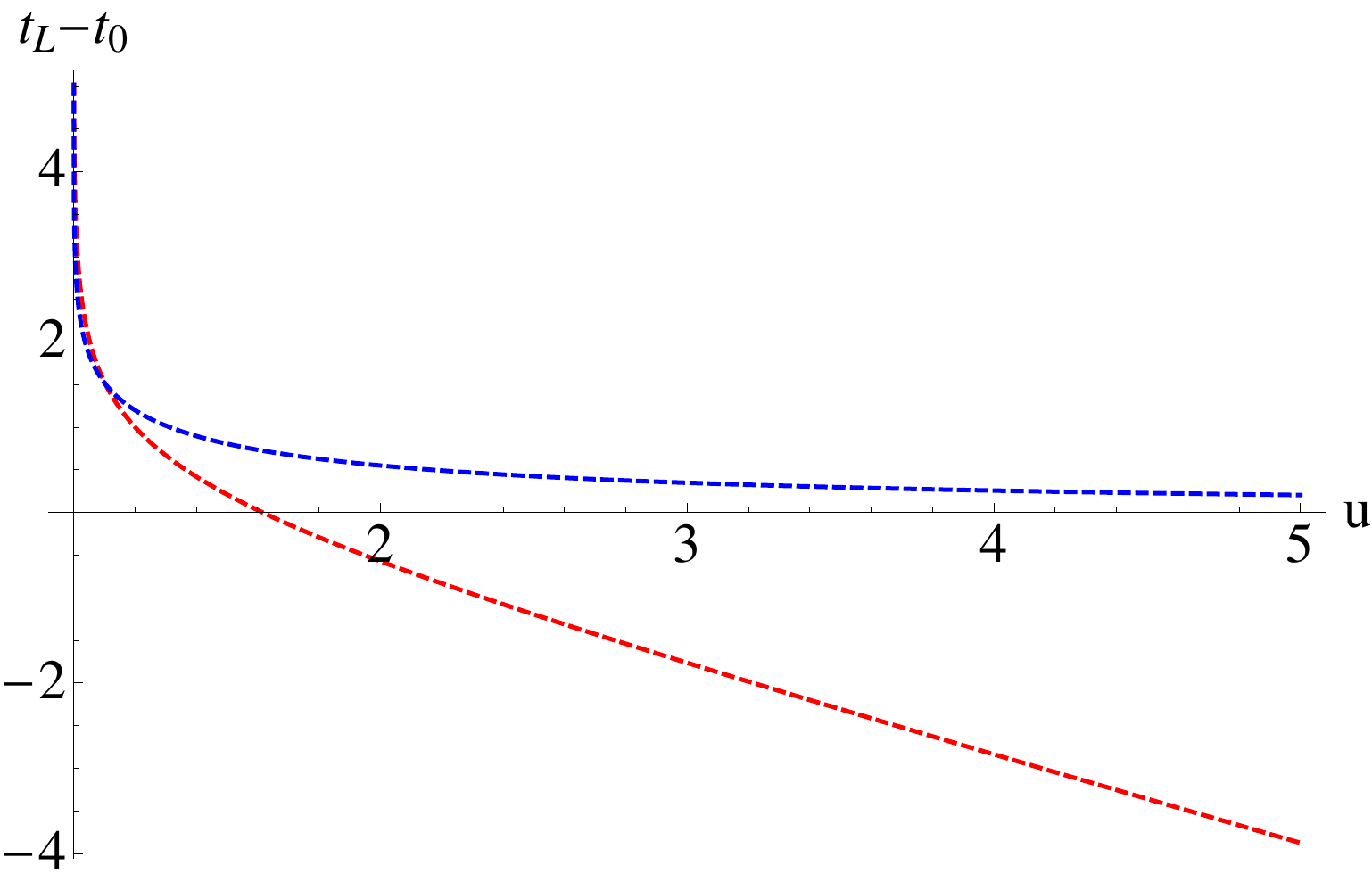}
\caption{Null geodesics inside the horizon: Lorentzian time $t_{\rm L}(u)-t_0$ along an ingoing radial null geodesic which starts from the boundary $u =0$ at $t = t_0$ for AdS$_3$-osm (dashed red lines) and BTZ with  $u_{\rm H}=u_*=1$ (dashed blue line). Euclidean time $t_{\rm E}=-\frac{\pi }{\sqrt{2}}$ for the AdS$_3$-osm case and $t_{\rm E}=-\frac{\pi }{2}$ for the BTZ case.}
\label{ngin}
\end{figure} 
%

\subsubsection{General geodesics}

We will now study the behaviour of general geodesics, focussing on the space-like ones, in the AdS$_3$-osm background in (\ref{osmetric}). Towards that end, the general radial geodesic equation can be written as: 
\begin{eqnarray}
&& \left(\frac{du}{d\lambda}\right)^2+ V_{\rm eff}(u) = 0 \ , \\
&& V_{\rm eff}(u) = \frac{u^2}{R^2} \left(1 - \frac{u^2}{u_*^2}\right) \left(-\kappa + \frac{L^2 u^2}{1+\frac{u^2}{u_*^2}} - \frac{P^2 u^2 }{1-\frac{u^4}{u_*^4}}\right) \ . \label{effV}
\end{eqnarray}
Here $V_{\rm eff}$ is the effective potential and $\lambda$ denotes the affine parameter along the geodesic. The parameter $\kappa=0,-1,+1$ for null, time-like and space-like geodesics, respectively. $P$ and $L$ are two constants of motion, corresponding to the two Killing vectors $\left( \partial / \partial \tau \right)$ and $\left( \partial / \partial x \right)$, respectively. We will thus associate an ``energy" with $P$ and a ``momentum" with $L$, henceforth. The effective potential at vanishing electric field is obtained by taking $u_* \to \infty$ limit, which yields:
\begin{equation}
V^{(0)}_{\rm eff}(u) = \frac{u^2}{R^2} \left(-\kappa + L^2 u^2 - P^2 u^2\right) \ . \label{adsV}
\end{equation} 
This is the result for pure-AdS background.

Let us make a few straightforward observations. Clearly, any geodesic enters the bulk, provided $\lim_{u \to 0} V_{\rm eff} < 0$. This condition translates to $L^2 < P^2$ for null geodesics, which can occur. For time-like geodesics, one gets $P^2 > \infty$, which means time-like geodesics never penetrate the bulk. Finally, for space-like geodesics, the condition is: $P^2 > - \infty$, which means space-like geodesics always penetrate the bulk. As a consistency check, one obtains the same conclusions by analyzing $V_{\rm eff}^{(0)}$, since AdS$_3$-osm asymptotically approaches AdS-geometry.

Let us now discuss the turning points of various geodesics, which we will denote by $u_{\rm c}$ and are given by solutions of $V_{\rm eff}(u_{\rm c}) = 0$. For null geodesics it is easy to check that no turning point exists. Thus all null geodesics that penetrate the bulk, and subsequently fall into the horizon.

For space-like geodesics, the possibilities are varied. Let us take the case when both $P = 0 = L$. The solutions of $V_{\rm eff} (u_{\rm c}) = 0$ gives $u_{\rm c} = 0, u_*$, of which $u_{\rm c} = 0$ can certainly not be counted as a turning point. One should, now, calculate the first derivative of the potential at $u_{\rm c}$ and check that it is indeed non-zero. A vanishing first derivative at $u_{\rm c}$ would imply an orbit, rather than a turning point. This is based on the intuition that in order for the geodesic to turn back from $u_{\rm c}$, one needs a non-vanishing ``force". It is straightforward to check that $V_{\rm eff}' (u_*) = 2 u_* > 0$. Thus, the event horizon $u_*$ is also a turning point and can be reached exactly by a geodesic. This is exactly similar to the BTZ-background.

Suppose now we assume $L=0$, $P \not = 0$. The only real-valued solution is:
\begin{eqnarray}
u_{\rm c} = \frac{u_*}{\sqrt{2}} \sqrt{u_*^2 P^2 + \sqrt{4 + u_*^2 P^2}} \ > u_* \ , 
\end{eqnarray}
and hence lies inside the horizon. Thus, there is no turning point in this case; the space-like geodesic falls inside the horizon and does not come back to the same boundary\cite{Hubeny:2012ry}. If we instead set $P = 0$ and $L \not =0$, then we get the following roots:
\begin{eqnarray}
u_{\rm c} = u_* \ , \ \frac{u_*}{\sqrt{u_*^2 L^2 -1}} \ .
\end{eqnarray}
Among the above, $u_*$ is an unstable orbit provided $u_* L = \sqrt{2}$ which obeys: $V_{\rm eff}'(u_*) = 0 $ and $V_{\rm eff}''(u_*) < 0 $. On the other hand, $u_{\rm c} = u_* / \sqrt{u_*^2 L^2 -1}$ is a turning point for $u_* L > \sqrt{2}$. For this root to be real and to lie outside the horizon one needs to impose $u_* L > \sqrt{2}$, which also satisfies $V_{\rm eff}'(u_{\rm c}) > 0$. Thus, in the regime $u_* L > \sqrt{2}$, there is only one true turning point: $u_{\rm c} = u_* / \sqrt{u_*^2 L^2 -1}$. Note that, by taking the limit $u_* L \to \sqrt{2}$, we immediately get $u_{\rm c} \to u_*$, and the two roots merge to become an unstable orbit. This is very similar to what one observes in a BTZ-geometry, see {\it e.g.}~\cite{Hubeny:2012ry}.

Finally, we can discuss the case in which both $P \not =0$, $L\not = 0$. Not surprisingly, for $P > L$, there is no turning point. For $P < L$, there are two roots within $0 < u_{\rm c} < u_*$. The smaller of these two roots correspond to the physical turning point of the space-like geodesic. As the ratio $(P / L)$ is raised, the two roots come closer to each other and finally coalesce at $P / L \approx \O(1)$. Beyond this, the roots disappear altogether. 
\begin{figure}[ht!]
\begin{center}
\includegraphics[width=0.8\textwidth]{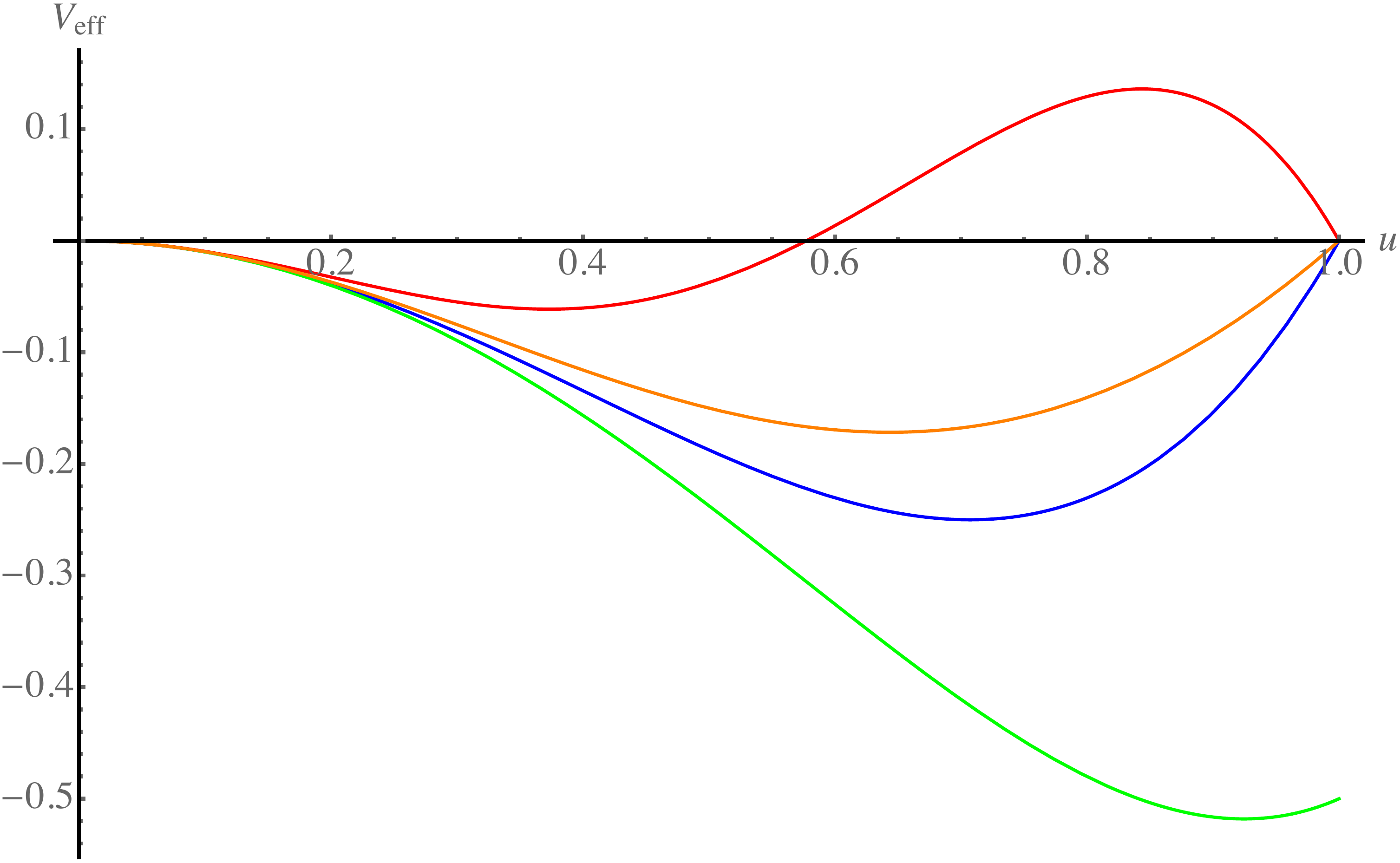}
\subfloat[Effective potential with $L < P$ and $L > P$.]
{\includegraphics[width=8.0cm]{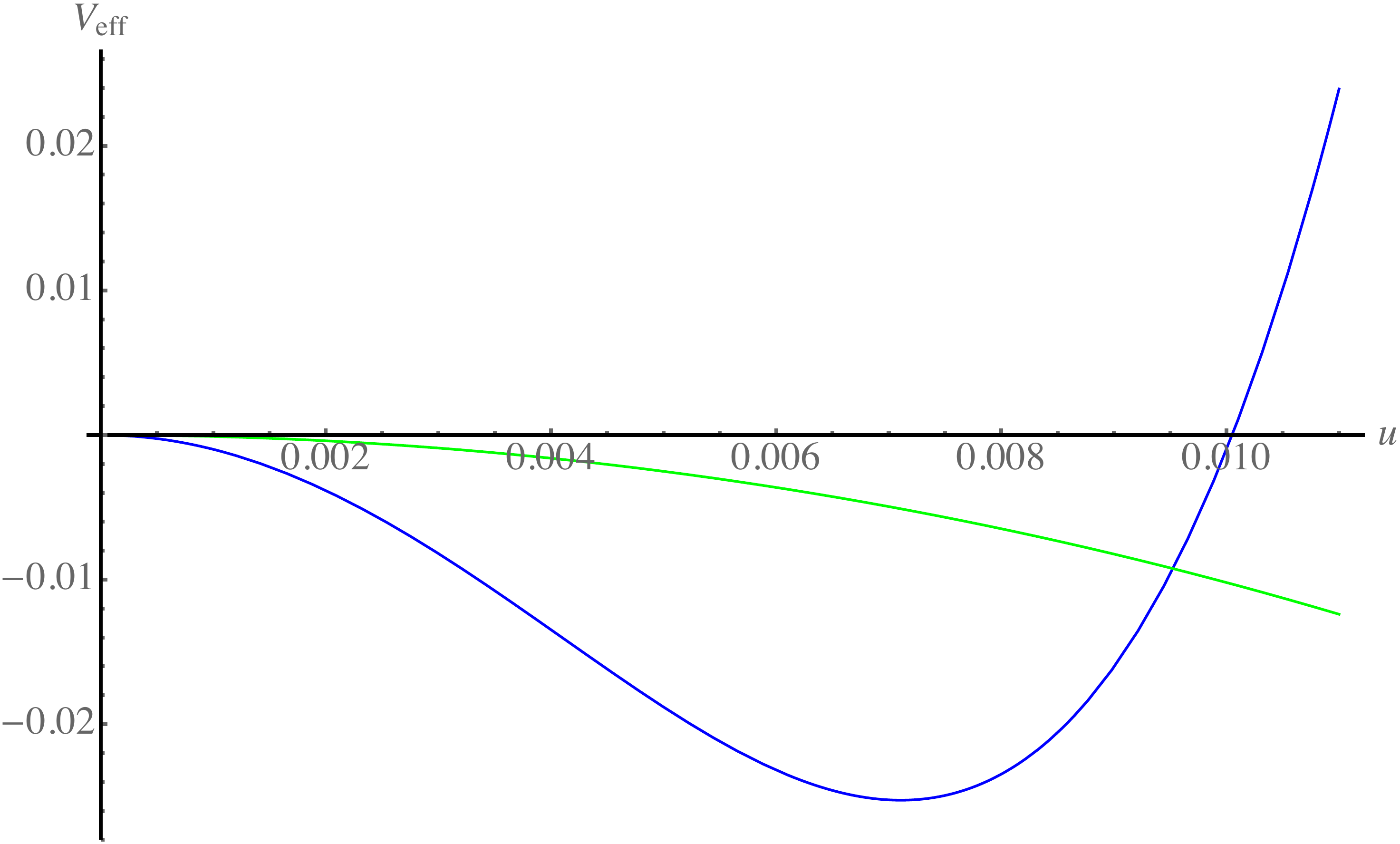}}
\subfloat[Effective potential with $L \sim P$.]
{\includegraphics[width=8.0cm]{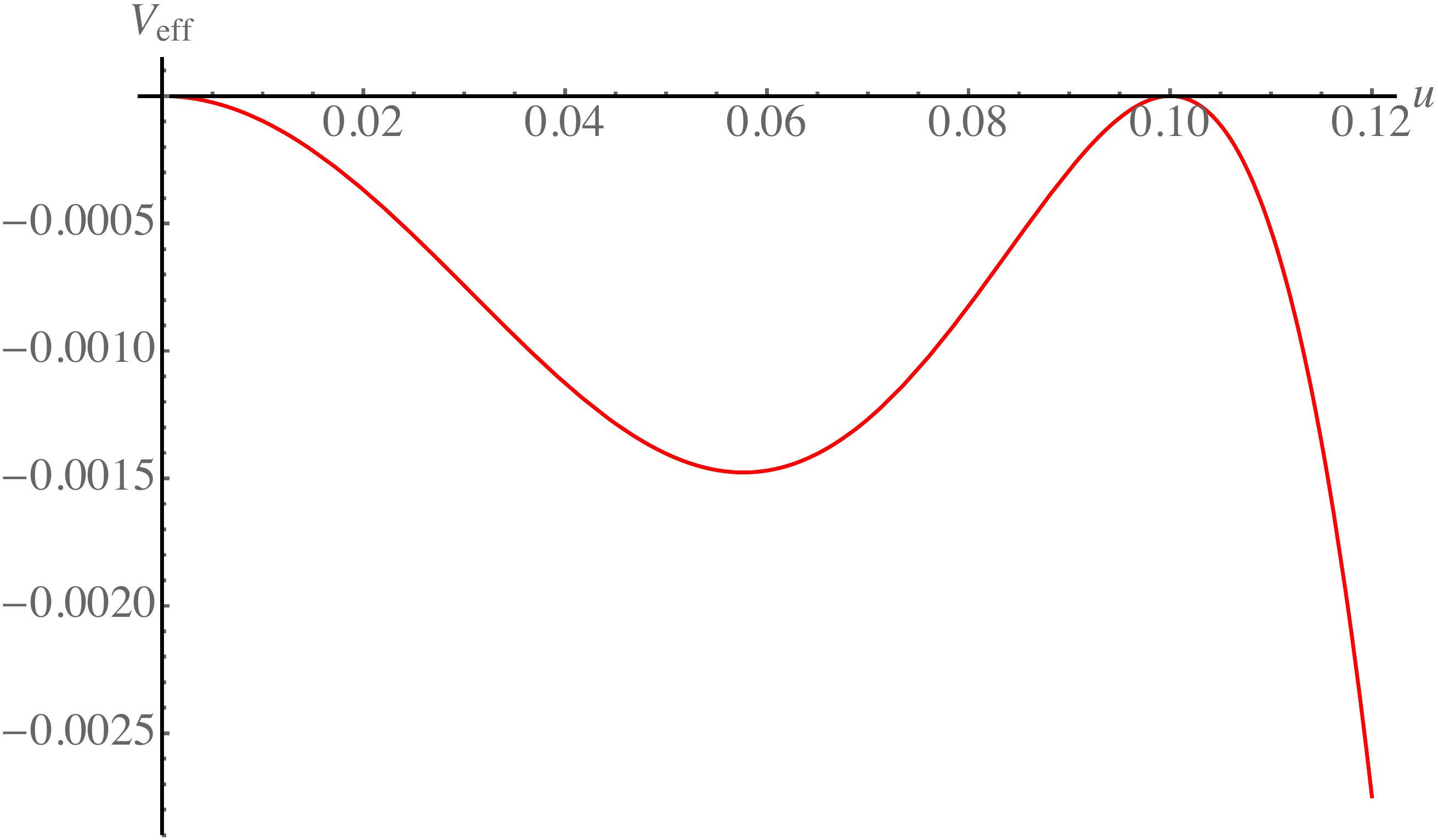}}
\caption{The top figure consists of a plot of the effective potential in (\ref{effV}) for spacelike geodesics with $L=0=P$ (blue), $L=0$, $P=1$ (green), $L=2$, $P=0$ (red) and $L=1$, $P=0$ (orange) respectively. In the bottom left figure, the green and blue curves correspond to $L < P$ and $L > P$, respectively. The latter evidently has a zero, but the former does not. The right figure depicts a fine-tuned value of $L \sim P$ such that the root just disappears. We have set $u_*=1$ in all of them.} \label{effpotenplot}
\end{center}
\end{figure}
To facilitate the discussion, our results are summarized in figure \ref{effpotenplot}.

Now, we want to study connected space-like geodesics at a fixed time-slice, which have two points anchored at the boundary. A fixed time-slice sets $P=0$. The separation between the points at the boundary is:
\begin{align}
l\equiv \Delta x(u=0)= & 2 \sqrt{\frac{1}{u_{\rm c}^2}+\frac{1}{u_*^2}} \int_0^{u_{\rm c}} \frac{udu}{\sqrt{\left(1-\frac{u^4}{u_*^4}\right) \left(1-\frac{ u^2}{u_{\rm c}^2}\right)}} \nonumber\\
= &2 u_{\rm c}^2 \sqrt{\frac{1}{u_{\rm c}^2}+\frac{1}{u_*^2}}  \, _3F_2\left(\frac{1}{2},\frac{1}{2},1;\frac{3}{4},\frac{5}{4};\frac{u_{\rm c}^4}{u_*^4}\right)\ .\label{uceqn}
\end{align}
Here $F$ denotes Apell function. The corresponding geodesic length is:
\begin{equation}
\L_{\rm E}(l)-\L_0(l) \equiv 2  \int_\epsilon^{u_{\rm c}} \frac{\sqrt{\left(1+\frac{u^2}{u_*^2}\right)}du}{u\sqrt{\left(1-\frac{u^2}{u_*^2}\right) \left(1-\frac{ u^2}{u_{\rm c}^2}\right)}} - 2  \int_\epsilon^{u_0} \frac{du}{u\sqrt{\left(1-\frac{ u^2}{u_0^2}\right)}} \ , \label{leqn}
\end{equation}
where $u_0 = l/2$. In the above expression, we have introduced a short-distance cut-off $\epsilon$ and subtracted off the geodesic length at vanishing electric field to regularize the integral. The resulting quantity is thus finite.

Following \cite{Fischler:2012ca, Kundu:2016dyk}, let us now consider two separate limits: (i) $u_{\rm c} \ll u_*$ and (ii) $u_{\rm c} \sim u_*$. Case (i), in which the geodesic stays mostly away from the infrared horizon, is equivalent to taking $l/u_* =l\sqrt{E} \ll 1$. In this limit, the equation (\ref{uceqn}) leads to:
\begin{equation}
u_{\rm c} = \frac{l}{2} \left(1-\frac{l^2}{8 u_*^2} + \O\left(l^4 u_*^4\right)\right)\ ,
\end{equation}
which gives:
\begin{equation}
\L_{\rm E}(l) - \L_0(l) \approx \frac{l^2}{2 u_*^2} = \frac{E l^2}{2} = \left(T_{\rm eff} l \right) ^2\pi^2 \ .
\end{equation}

On the other hand, when $u_{\rm c} \sim u_*$, both $l$ and $\L_{\rm E}(l) - \L_0(l)$ diverges. Nevertheless, one can check that, as $u_{\rm c} \rightarrow u_*$, the following quantity
\begin{equation}
\L_{\rm E}(l) - \L_0(l)-\frac{\sqrt{2} l}{u_*} = \left[c + 2 \ln \left(\frac{u_*}{l}\right) \right] \ ,
\end{equation}
remains finite, where
\begin{equation}
c=\int_0^1 \left(\frac{2 \sqrt{x^2+1}}{x (1-x^2 )}-\frac{4 x}{\sqrt{1-x^4} \sqrt{1-x^2}}-\frac{2}{x}\right) \, dx \approx -0.376\ .
\end{equation}
Hence we get
\begin{equation}
\L_{\rm E}(l) - \L_0(l) = \frac{\sqrt{2} l}{u_*} + \left[c + 2\ln \left(\frac{u_* }{l}\right)\right] = \left (2 \pi T_{\rm eff}l + c + 2 \ln \left(\frac{1}{\sqrt{2}\pi T_{\rm eff}l} \right) \right) \ .
\end{equation}
It is natural to interpret this quantity as an equal-time two-point function of quantum fluctuations that live on the probe brane:
\begin{equation}
\langle \J(t,0)\J(t,l)\rangle_{\rm c} \equiv 1- e^{-\Delta(\L_{\rm E}(l)-\L_0(l))}\ ,
\end{equation}
where $\Delta$ is a dimensionless constant.
\begin{figure}[ht!]
\centering
\includegraphics[width=0.8\textwidth]{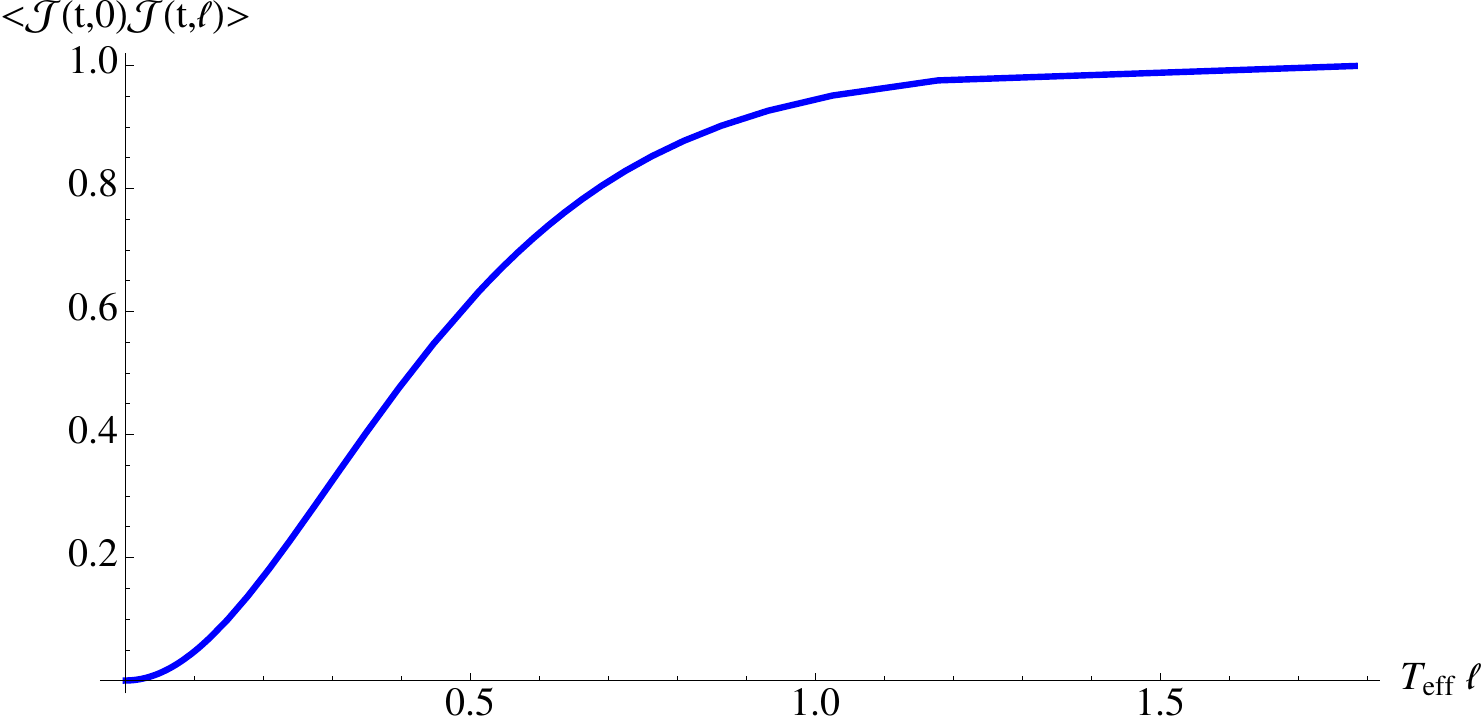}
\caption{Variation of $\langle \J(t,0)\J(t,l)\rangle_{\rm c}$ with $T_{\rm eff} l=\frac{E^{1/2}l}{\sqrt{2}\pi}$ for $\Delta=1$.}
\end{figure}

Before concluding this section, let us offer some more comments, specially involving the Gao-Wald result in \cite{Gao:2000ga}. Let us explicitly write the general geodesic equations using the metric in (\ref{tautdef}). Using (\ref{cons}) and (\ref{defgeoex}), one gets:
\begin{eqnarray}
&& \frac{1}{u^2} \left( 1 - \frac{u^4}{u_*^4} \right) \dot{\tau} = P  \ , \quad \left( \frac{1}{u^2} + \frac{1}{u_*^2} \right) \dot{x} = L \ , \\
&& \dot{u}^2 + V_{\rm eff} = 0 \ , \quad V_{\rm eff} = u^2 \left(1 - \frac{u^2}{u_*^2} \right) \left[ - \kappa - \frac{P^2 u^2}{1 - \frac{u^4}{u_*^4}} + \frac{L^2 u^2}{1 + \frac{u^2}{u_*^2}} \right] \ .
\end{eqnarray}
Suppose we want to focus on the geodesics whose end points are located at the conformal boundary at $u = 0$. The temporal and the spatial distances between the two points of the geodesic is given by
\begin{eqnarray}
\Delta \tau & = & 2 \varepsilon \int_{u_{\rm c}}^{0} \frac{u^2 du}{\left( 1 - \frac{u^4}{u_*^4}\right) \sqrt{- V_{\rm eff}}} \ , \\
\Delta x = & = & 2 \int_{u_{\rm c}}^{0} \frac{u^2 du}{\left( 1 + \frac{u^2}{u_*^2}\right) \sqrt{- V_{\rm eff}}} \ , \quad \varepsilon = \frac{P}{L} \ .
\end{eqnarray}
By examining the integrands in the above expressions it trivially follows that, similar to what was observed in \cite{Hubeny:2012ry}, $\Delta \tau \ge \varepsilon^{-1} \Delta x$. This is in clear coherence with the Gao-Wald inequality of \cite{Gao:2000ga}, {\it i.e.}~$\Delta \tau \ge \Delta x$, provided $\varepsilon^{-1} > 1$. The last inequality is needed for the geodesic to have a turning point, and hence it is consistent. Let us also remark that, even though we will discuss later issues with ``energy conditions" associated with the osm, the Gao-Wald result only relies on a null Ricci condition that nonetheless is preserved by the osm.

\begin{figure}[ht!]
\centering
\includegraphics[width=0.8\textwidth]{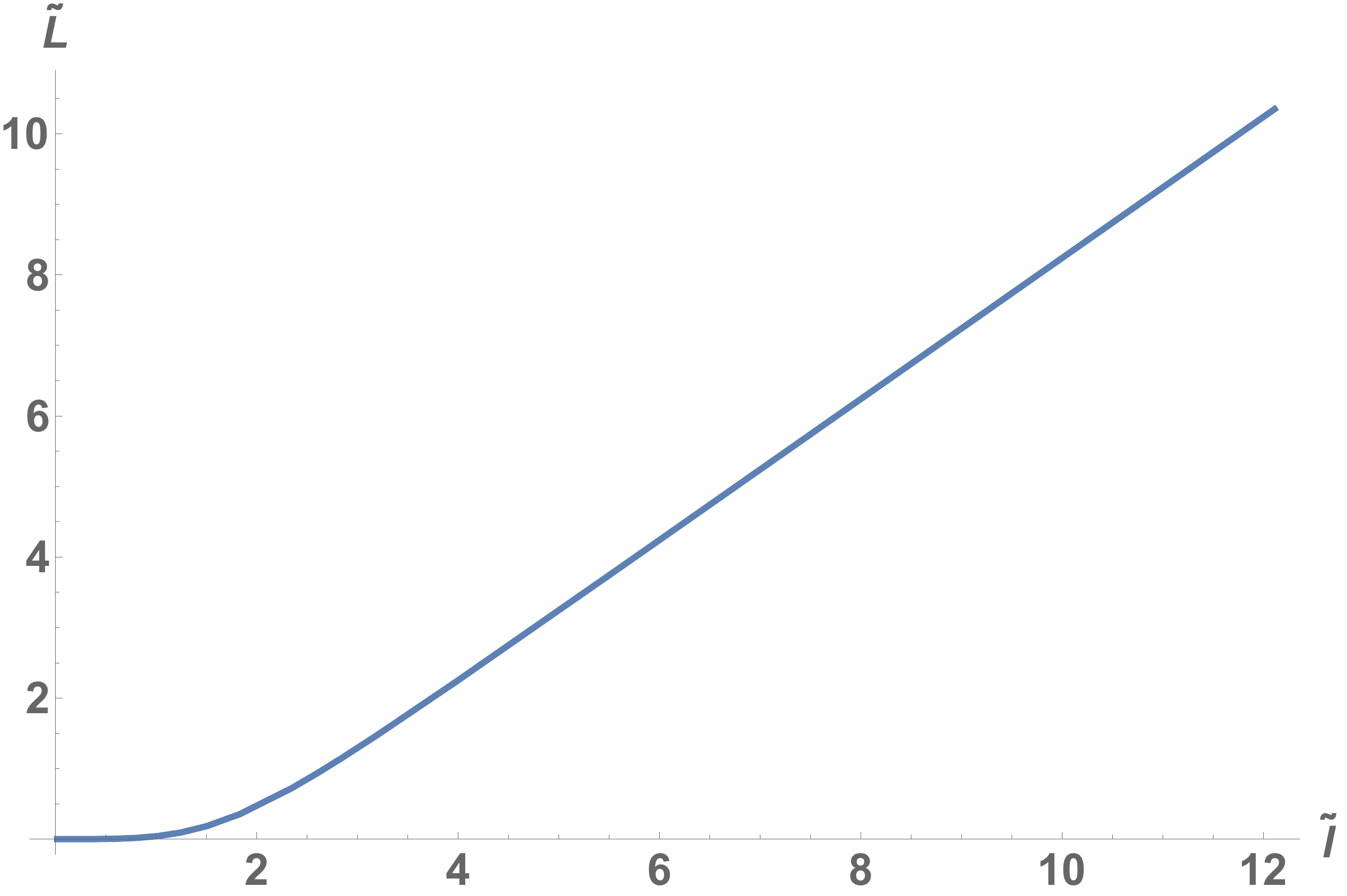}
\caption{Variation of the renormalized geodesic length in the $\tilde{\S}$-geometry, demonstrated with respect to the boundary separation length $\tilde{l}$. The latter is measured in units of the effective temperature.}
\label{corrtilde}
\end{figure}

So far, our discussion has been based on $\S$, instead of $\tilde{\S}$. A similar computation is straightforward for $\tilde{\S}$-background. We refrain from providing the details of this, except the following ones:
\begin{eqnarray}
\tilde{l} &  = & 2 u_{\rm c} \, _3F_2\left(\frac{1}{2},\frac{1}{2},1;\frac{3}{4},\frac{5}{4};\frac{u_{\rm c}^4}{u_*^4}\right) \ , \\
\tilde{L} & = & \int_{\epsilon}^{u_{\rm c}} \frac{du}{u} \frac{2}{\sqrt{\left( 1 - \frac{u^4}{u_*^4}\right) \left( 1- \frac{u^2}{u_{\rm c}^2}\right)  }} - \int_{\epsilon}^{u_{\rm c}} \frac{du}{u} \frac{2}{\sqrt{\left( 1- \frac{u^2}{u_{\rm c}^2}\right) }} \ ,
\end{eqnarray}
where $\tilde{l}$ and $\tilde{L}$ denote the boundary separation and the length of the geodesic in $\tilde{\S}$-geometry, respectively. Finally, we present the behaviour of the length of a spacelike geodesic in figure \ref{corrtilde}.
Qualitatively, this behaviour is similar to what is observed in the pure osm-background. Similarly, the Gao-Wald conclusion also holds for the conformal osm-geometry, which, as we will show momentarily, satisfies the null Ricci condition.

\subsubsection{General geodesics in higher dimensions}

In this section, we will discuss features of the effective potential $V_{\rm eff}$ in an AdS$_4$ and an AdS$_5$-background. The corresponding open string metrics are given by:
\begin{eqnarray}
ds_{(4)}^2 & = & - \frac{1 - E^2 u^4}{u^2} d\tau^2 + \frac{du^2}{u^2 \left(1 - E^2 u^4 \right)} + \frac{1}{u^2} d\vec{x}^2 \ , \label{4osm} \\
ds_{(5)}^2 & = &  - \frac{1 - E^2 u^4}{u^2} d\tau^2 + \frac{du^2}{u^2 \left(1 - E^3 u^6 \right)} \nonumber \\
& + & \frac{1 + E u^2}{u^2 + E u^4 + E^2 u^6} dx^2 + \frac{1}{u^2} dx_{\perp}^2  \ . \label{5osm}
\end{eqnarray}
We also define
\begin{eqnarray}
&& \tilde{\S}_{(4)} = \Omega_{(4)} \S_{(4)} \ , \quad  \tilde{\S}_{(5)} = \Omega_{(5)} \S_{(5)} \ , \\
&& \Omega_{(4)} = 1 \ , \quad  \Omega_{(5)} = \left( 1 + \frac{u^2}{u_*^2} + \frac{u^4}{u_*^4} \right)^{1/3} \left( 1 + \frac{u^2}{u_*^2} \right)^{-1/3} \ . \label{omega45}
\end{eqnarray}
Note that, in AdS$_4$, the isotropy in the $\{x, y\}$-plane is not broken by the electric field. In AdS$_5$, it is broken and we denote the directions perpendicular to the electric field by $x_\perp$. Thus, while considering spatial geodesics, in the later case one can consider two points parallel or perpendicular to the electric field. As before we can write down the geodesic equation with an effective potential. These effective potentials are respectively given by
\begin{eqnarray}
&& V_{\rm eff}^{(4)} = u^2 \left( 1 - \frac{u^4}{u_*^4} \right) \left[ - \kappa + L^2 u^2 - \frac{P^2 u^2}{1 - \frac{u^4}{u_*^4}}\right] \ ,  \quad L^2 = L_x^2 + L_y^2 \ , \label{veff4} \\
&& V_{{\rm eff}(||)}^{(5)} = u^2 \left( 1- \frac{u^6}{u_*^6}\right) \left[ - \kappa + L_{||}^2 \frac{u^2 \left( u^4 + u_*^2 u^2 + u_*^4\right) } {u_*^2 \left( u^2 + u_*^2 \right) } - \frac{P^2 u^2}{1 - \frac{u^4}{u_*^4}} \right]  \ , \label{veff5pa} \\
&& V_{{\rm eff}(\perp)}^{(5)} = u^2 \left( 1- \frac{u^6}{u_*^6}\right) \left[ - \kappa + L_{\perp}^2 u^2 - \frac{P^2 u^2}{1 - \frac{u^4}{u_*^4}} \right]  \ , \quad L_{\perp}^2 = L_y^2 + L_z^2 \ . \label{veff5pe}
\end{eqnarray}
with $u_* = E^{-1/2}$.

From (\ref{veff4}), it is clear that spacelike geodesics will penetrate the geometry for any given energy. The structure is exactly the same as the asymptotic AdS$_3$-example for both cases when $P = 0 = L$ and $L=0$, $P\not = 0$. In the case $L \not = 0$, $P=0$ the two roots are:
\begin{eqnarray}
u_{\rm c} = u_* \ , \ \frac{1}{L} \ .
\end{eqnarray}
The above two roots are turning points provided $ u_* L < 1$ and $u_* L > 1$, respectively. If $u_* L =1$, both roots become stable orbits. In the most general case, for $P>L$ there is no turning point, for $P<L$ there are two roots and as the ratio $(P/L)$ is increased they coalesce at $P / L \approx \O(1)$ and disappear after this. It is also straightforward to verify that all qualitative features remain the same with the effective potentials in (\ref{veff5pa}) and (\ref{veff5pe}).

\subsection{Kruskal extension and Penrose diagram}

Given the geometry in (\ref{osmetric}) in asymptotically AdS$_3$, or the geometry in (\ref{4osm}) in asymptotically AdS$_4$ background, we can work out the corresponding Kruskal extension of those, and subsequently the Penrose diagrams. In this section we will discuss them, relegating the details of the calculations -- that mostly involves a chain of coordinate transformations -- to a couple of appendices: Appendix \ref{sec:KrusBTZ}, in which the standard Kruskal extension of BTZ-geometry is reviewed, then appendix \ref{sec:osmKrus3} and \ref{sec:osmKrus4} where we discuss the Kruskal extension of the open string metric in asymptotically AdS$_3$ and AdS$_4$ backgrounds, respectively. 
\begin{figure}[ht]
\begin{center}
\subfloat[Penrose diagram for AdS$_3$-osm geometry.]
{\includegraphics[width=7.9cm]{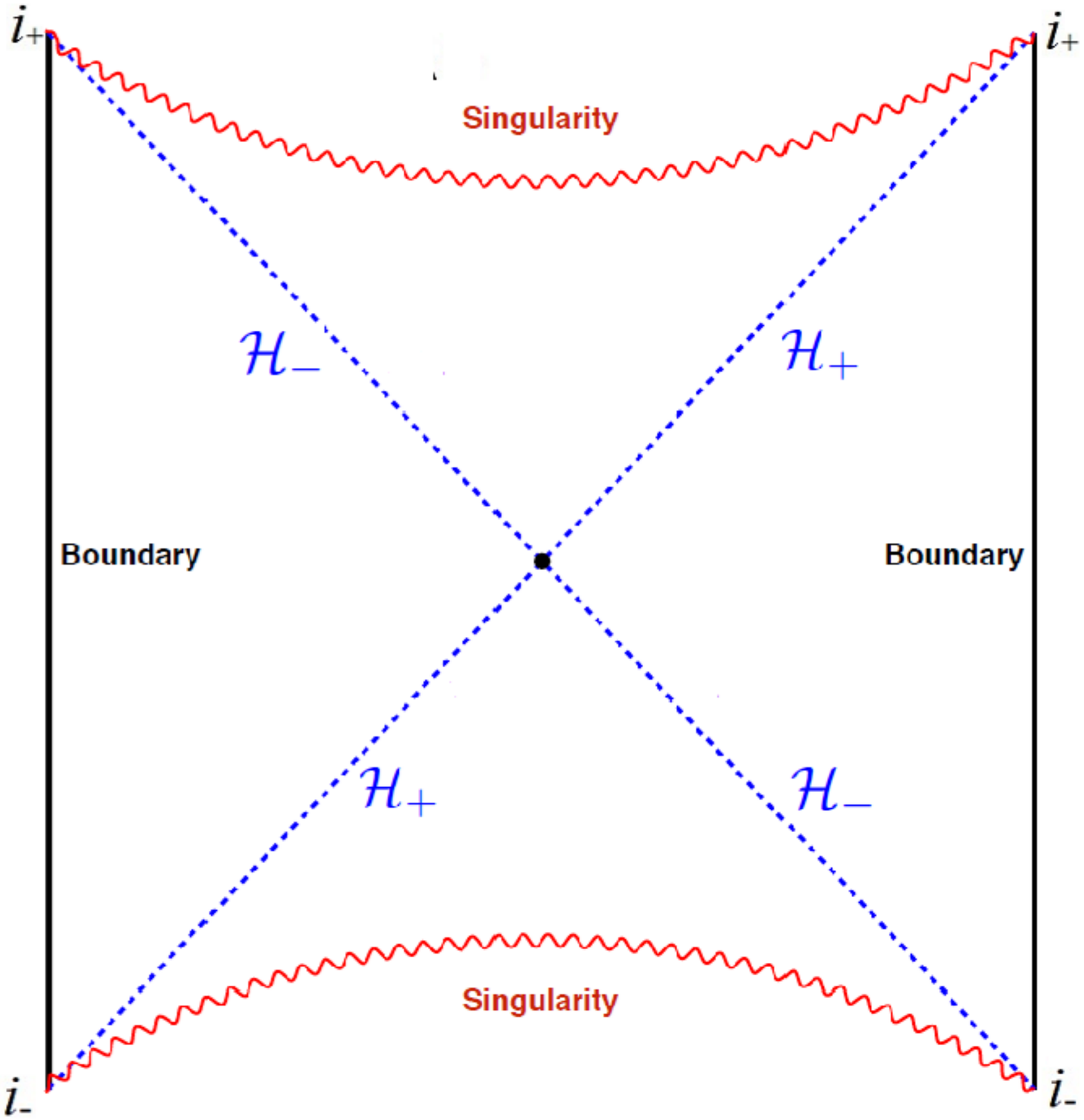}}
\subfloat[Penrose diagram for AdS$_4$-osm geometry.]
{\includegraphics[width=7.9cm]{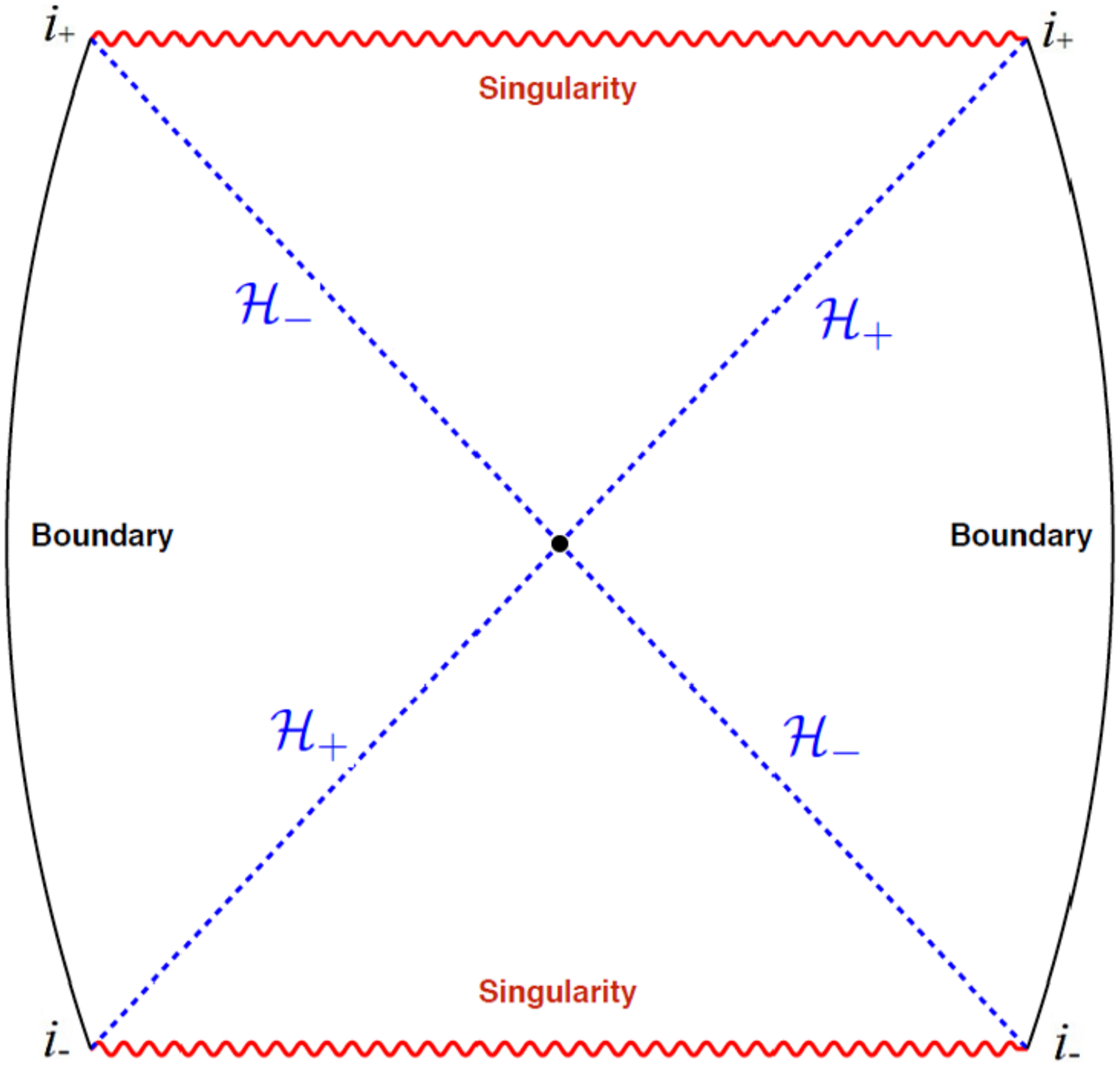}}
\caption{Penrose diagrams for osm-geometry. The black dot in both the diagrams correspond to the bifurcation surface, {\it i.e.}~the intersection point of the future and the past horizons. Both these diagrams are qualitatively similar, up to a conformal transformation.}
\label{pen34}
\end{center}
\end{figure}
The resulting diagrams are shown in fig.~\ref{pen34}.

First, let us note the qualitative similarities to an eternal Schwarzschild black hole which, in turn, are responsible for the effective ``thermal" description of the system. There is a technical difference though: It is known that for BTZ black hole, the corresponding Penrose can be drawn as a perfect square, {\it i.e.}~null rays originating at $T=0$ (the Kruskal time, see {\it e.g.}~equation (\ref{btzTX})) from the causally disconnected AdS-boundaries meet precisely at the location of the singularity. In this case, however, either the singularity bends inwards (while the conformal boundary is represented by a straight vertical line), or the conformal boundary bends outwards (while the singularity is represented by a straight horizontal line). This means that two null rays originating at the analogue of $T=0$ from the two different AdS-boundaries will encounter the singularity before they meet each other. This feature is similar to the Penrose diagram of AdS-BH geometries in higher dimensions. Finally, let us comment on the apparent distinction of the AdS$_3$-osm and the AdS$_4$-osm Penrose diagrams. We emphasize here, as also discussed in appendix \ref{sec:osmKrus4}, that by a conformal transformation we can represent the conformal boundary as a straight vertical line, at the cost of bowing the singularity inwards. Thus, qualitatively, both Penrose diagram for AdS$_3$-osm and AdS$_4$-osm are similar.

Let us also note that, even though Kruskal extension seems a natural mathematical step to follow, the physical origin of doing such is not as clear in the case of an osm-geometry. In a given geometry various Kruskal patches are related by local diffeomorphisms, and thus all of them satisfy the same Einstein equations. A particular osm-geometry does not originate from directly solving any equation. Thus, Kruskal extension of a given osm-geometry should perhaps be practiced with some reservations.

In view of the above, one might wonder about the following: Suppose we begin with the Kruskally extended AdS$_3$-BH geometry. Now, is there a probe brane configuration that will produce the previously-obtained Kruskally-extended osm patch? The answer to this question seems to be no. Thus, even though we point out the otherwise similarities of the Kruskal patch and the subsequent Penrose diagram of the osm, we emphasize that this is subtle.

\section{A Note on Energy/Ricci Conditions}

Let us now discuss, supposing that the open string geometry could emerge from Einstein-gravity sourced with appropriate matter field, the nature of the corresponding stress tensor. As before, we will begin the discussion with asymptotically AdS$_3$-background in (\ref{osmetric}), and subsequently comment on the higher dimensional analogues in (\ref{4osm}) and (\ref{5osm}). To start with, let us choose the notation: we use $\G_{\mu\nu}$ to denote the corresponding Einstein-tensor evaluated from the given open string metric. The {\it equation} we pretend solving is the following:
\begin{eqnarray}
\G_{\mu\nu} + \Lambda g_{\mu\nu} = \T_{\mu\nu} \ , \label{einpretend}
\end{eqnarray}
where $\Lambda = - d(d-1) / 2$ is the cosmological constant in asymptotically AdS$_{d+1}$-background, $G_{\mu\nu}$ is the open string metric and $\T_{\mu\nu}$ is the putative matter field. We will investigate the following energy conditions:\\

\noindent {\bf (i) Null Energy Condition (NEC):} This implies that for every future pointing null vector, the matter density observed by the corresponding observer is non-negative. For a given $\T_{\mu\nu}$ and any null vector $n^\mu$, the null energy condition imposes: $\T_{\mu\nu} n^\mu n^\nu \ge 0$. 

\noindent {\bf (ii) Weak Energy Condition (WEC):} It implies that for every future pointing timelike vector, the matter density observed by the corresponding observer is non-negative. For a given $\T_{\mu\nu}$ and any timelike vector $t^\mu$, the null energy condition imposes: $\T_{\mu\nu} t^\mu t^\nu \ge 0$. 

\noindent {\bf (iii) Strong Energy Condition (SEC):} This condition originally is imposed on the Ricci tensor and makes a reference to the Einstein field equation to be recast in terms of the stress-tensor. By virtue of Raychaudhuri equation, the timelike convergence condition becomes: $R_{\mu\nu} t^\mu t^\nu \ge 0$.\footnote{Note that, the Raychaudhuri equation, in the absence of shear and rotation, can be written as: $\dot{\Theta} + \frac{1}{3} \Theta^2 = - R_{\mu\nu} t^\mu t^\nu$, where $\Theta$ is the expansion and $t^\mu$ denotes a timelike unit vector field. For gravity to be an attractive force, the RHS of the equation should remain negative, and hence a condition on the Ricci tensor follows.} Using (\ref{einpretend}) this condition is equivalent to:
\begin{eqnarray}
\left[ \left(\T_{\mu\nu} - \frac{1}{d-1} G_{\mu\nu} \T \right) + G_{\mu\nu} \Lambda \frac{2}{d-1} \right] t^\mu t^\nu \ge 0 \ , \quad \Lambda = - \frac{d(d-1)}{2} \ ,
\end{eqnarray}
for asymptotically AdS$_{d+1}$-background.

To make the discussion self-contained, let us briefly remark when $\T_{\mu\nu}$ corresponds to an ideal fluid in a $(D+1)$-dimensional Minkowski background and is given by
\begin{eqnarray}
\T_{\mu\nu} = \left(\rho + p \right) u_\mu u_\nu + p \ \eta_{\mu\nu} \ ,
\end{eqnarray}
where $\rho$ is density and $p$ is the pressure. In this case, Null Energy Condition implies $\rho \ge 0$, Weak Energy Condition implies $\rho \ge 0$ and $\rho + p \ge 0$ and Strong Energy Condition implies $\rho + p \ge 0$ and $\rho + D p \ge 0$.

Let us now discuss the backgrounds in (\ref{osmetric}), (\ref{4osm}) and (\ref{5osm}). For subsequent discussions, let us pick the timelike and null vectors:
\begin{eqnarray}
t^\mu = \left( \frac{1}{\sqrt{|G_{\tau\tau}|}}, \ldots \right) \ , \quad n^\mu = \left( \frac{1}{\sqrt{|G_{\tau\tau}|}}, \frac{1}{\sqrt{G_{uu}}} \ldots \right) \ .
\end{eqnarray}
Evaluated with (\ref{osmetric}), the various components are:
\begin{eqnarray}
&& \T_{\tau\tau} = \frac{ 2 u^2 \left(u_*^2 -u^2 \right)}{ u_*^4 \left(u_*^2+u^2 \right)} \ , \quad \T_{uu} = - \frac{2u_*^4}{\left(u_*^2 - u^2\right) \left(u_*^2 + u^2 \right)^2} \ , \\
&& \T_{xx} = - \frac{2u^2 \left(2u_*^2 + u^2 \right)}{u_*^4 \left(u_*^2 + u^2 \right)} \ . 
\end{eqnarray}
A straightforward calculation shows that 
\begin{eqnarray}
\T_{\mu\nu} t^\mu t^\nu = \frac{2u^4}{\left(u_*^2  +u^2 \right)^2} \ge 0 \ ,
\end{eqnarray}
and hence WEC is satisfied. Similarly, the SEC evaluates to: $R_{\mu\nu} t^\mu t^\nu = 2 \left( 1- \frac{u^2}{u_*^2}\right) \ge 0$, and is also satisfied. The NEC, on the other hand, yields:
\begin{eqnarray}
\T_{\mu\nu} n^\mu n^\nu = - \frac{2u^2 \left(u_*^2 - u^2 \right)}{ \left(u_*^2 + u^2 \right)^2} < 0 \ , \label{necvio}
\end{eqnarray}
and is violated. Note that in the limit $u \to 0$, (\ref{necvio}) approaches zero from negative; thus, the pure AdS-limit is recovered. Summarizing, the AdS$_3$-osm satisfies the WEC and SEC, but violates NEC.\footnote{Although we do not explicitly discuss it in details, it can be checked that the Dominant Energy Condition is satisfied in this case, while the Null Dominant Energy Condition is violated.}

Let us now move up one dimension. In AdS$_4$-osm, with a similar choice for the timelike and the null vector, it can be easily checked that the WEC evaluates to: $\T_{\mu\nu} t^\mu t^\nu = - u^4 / u_*^4 < 0$ and is thus violated. On the other hand, the NEC evaluates to $\T_{\mu\nu} n^\mu n^\nu = 0$ and the SEC yields: $R_{\mu\nu} t^\mu t^\nu = 3 - u^4/u_*^4 > 0$. Hence both NEC and SEC are satisfied. Subsequently, it can also be shown that the Dominant Energy Condition and the Null Dominant Energy Condition is satisfied. We will refrain from presenting further details, but comment that a similar conclusion is reached for AdS$_5$-osm background as well. We summarize this section by stating that, either the WEC or the NEC is violated in these osm-geometries, however, the other energy conditions may continue to hold.

Let us end this section with a final remark. In the above we have assumed a {\it trivial profile} for the probe embedding. A natural question is whether a probe with {\it sufficient bending} may lead to an osm that can subsequently be viewed to satisfy all the energy conditions. With a simple, instructive example we will demonstrate that this is not the case. However, we will not be rigorous. Suppose we start in an asymptotically AdS$_4$ geometry and let us assume a representative brane embedding that spans $\{t, u, x^1\}$-directions and has a non-trivial profile, characterized by the function $x^2(u)$. Of course, we are shying away from the full ten-dimensional details and that is part of the lack of rigour in our example. With this, it can be checked that:
\begin{eqnarray}
&& \T_{\mu\nu} t^\mu t^\nu = \frac{u_*^4 s^2 + 2 u_*^2 s^2 u^2 + \left( 2 + s^2\right) u^4}{\left( 1 + s^2 \right) \left( u_*^2 + u^2\right)^2 } \ge 0 \ , \quad \implies \quad {\rm WEC \ satisfied} \ , \\
&& \T_{\mu\nu} n^\mu n^\nu = - 2 \frac{u^2 \left( u_*^2 - u^2 \right) }{\left( 1 + s^2 \right) \left( u_*^2 + u^2\right)^2 } \le 0 \ , \quad \implies \quad {\rm NEC \ violated} \ , \\
&& {\rm SEC} \quad \implies \quad 2 \frac{\left(u_*^2 - u^2 \right)}{u_*^2 \left( 1 + s^2 \right) } \ge 0 \ , \quad {\rm hence\ satisfied} \ , \\
&& {\rm where} \quad s(u) = \frac{d x^2}{du} \ .
\end{eqnarray}
Thus, the resulting three-dimensional osm indeed behaves much like the AdS$_3$-osm that we have discussed above: it satisfies the WEC, but violates NEC. 

The astute reader would notice that so far we have dealt with the open string metric. In view of (\ref{canokin}) and (\ref{confosm}), one may ask of similar questions for the conformal metric $\tilde{\S}$. Now we comment on the results obtained for the corresponding ``energy-momentum" tensor, denoted by $\tilde{\T}$. With the help of (\ref{confosm3}), for asymptotically AdS$_3$, it is easy to check that:
\begin{eqnarray}
&& \tilde{\T}_{\mu\nu} t^\mu t^\nu = - \frac{u^4}{u_*^4} < 0 \quad \implies \quad {\rm WEC \, \, violated} \ , \\
&&  \tilde{\T}_{\mu\nu} n^\mu n^\nu = \frac{u^2}{u_*^4} \left( u_*^2 - u^2 \right) > 0 \quad \implies \quad {\rm NEC \, \, satisfied} \ , \\
&& {\rm SEC} \quad \implies \quad 2 + \frac{u^2}{u_*^2} > 0 \ , \quad {\rm hence \, \, satisfied} \ .
\end{eqnarray}
For asymptotically AdS$_4$, $\Omega =1$ and thus all conclusions remain same as in the osm-background. Furthermore, using (\ref{omega45}), it is straightforward to check that, for asymptotically AdS$_5$-background, WEC is violated, while both NEC and SEC are satisfied, corresponding to the conformal osm geometry. In brief, the the conformal osm always violates the WEC. This, in turn, implies that there is no area-increase theorem for the osm horizon area, and consequently, we cannot identify this area with thermal entropy. Finally, we should be careful in placing physical importance on the energy conditions, {\it e.g.}~it is known that a single scalar field with a non-trivial potential, which would otherwise be considered as innocuous, can violate SEC.

\section{Causal Holographic Observables}

An interesting aspect of gauge-string duality is to understand how the bulk geometry can be reconstructed from the boundary field theory data. In recent years, intriguing suggestions have been made towards this and indicative connections to quantum information theory established. The most popular of these is the entanglement entropy proposal in the context of holography\cite{Ryu:2006bv}. In our case, this proposal simply boils down to the computation of co-dimension two extremal area (simply space-like geodesics for AdS$_3$) surfaces. However, the Ryu-Takayanagi formula is widely believed to hold for Einstein gravity, and not necessarily for an effective or emergent geometry.

A more inherently geometric quantity is the Causal Holographic Information (CHI), introduced and analyzed in \cite{Hubeny:2012wa}. As long as there is a well-defined causal structure, the CHI is a well-defined concept. We will, in this section, offer some comments on the CHI evaluated on the osm backgrounds. Following \cite{Hubeny:2012wa}, given a spacelike region $\A$ in the boundary, one naturally associates a domain of dependence denoted by $\lozenge_{\A}$. The bulk causal wedge, denoted by $\blacklozenge_\A$, is then defined as the intersection of causal past and causal future of $\lozenge_{\A}$ itself. The past and future boundaries of the causal wedge intersect at a co-dimension $2$ surface, known as the causal information surface and is denoted by $\Xi_{\A}$.

Given an asymptotically AdS$_3$-background, we can choose $\A$ to be an interval of width $2a$. This, along with the subsequent choice of $\lozenge_\A$ can then be written as:
\begin{eqnarray}
&& \A = \left\{ \left( \tau, x\right) \big | \tau = 0, |x| \le a \right \} \ , \\
&& \lozenge_\A = \left\{ \left( \tau, x\right) \big | |\tau| + x \le a, x\in [0,a] \right\} \cup  \left\{ \left( \tau, x\right) \big | |\tau| - x \le a, x\in [-a,0] \right\} \ . 
\end{eqnarray}
To construct the bulk causal wedge, we need to analyze null geodesics in the corresponding geometry.

\begin{figure}[ht!]
\centering
\includegraphics[width=0.8\textwidth]{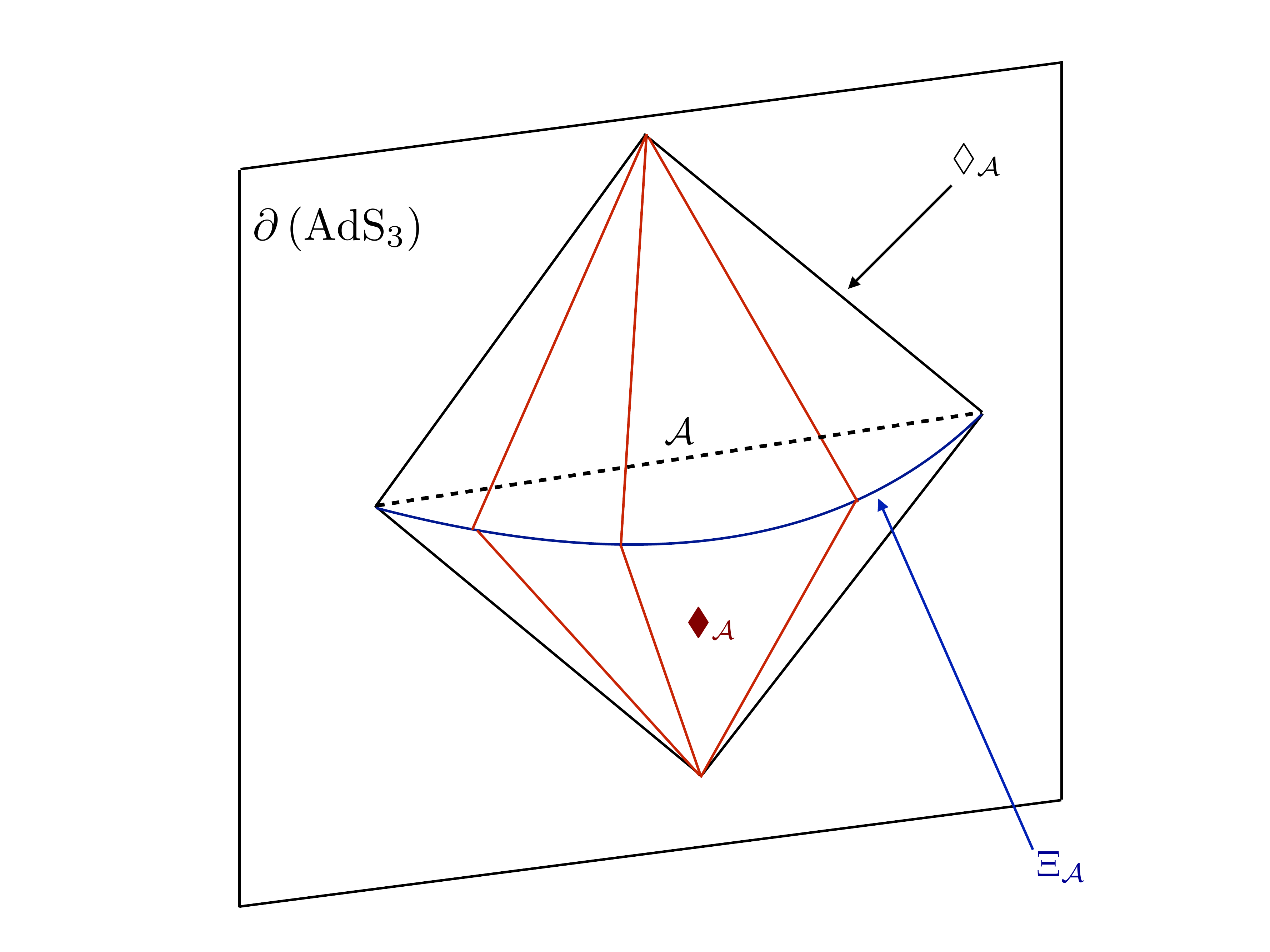}
\caption{The typical picture of the causal holographic information. This is very similar to what one obtains in {\it e.g.}~a BTZ-background.}
\label{chipic}
\end{figure}

With this in mind, we use the equations in (\ref{cons}) and (\ref{defgeoex}) for null geodesics in the background (\ref{osmetric}) we get
\begin{eqnarray}
&& \frac{1}{u^2}\left(1-\frac{u^4}{u_*^4}\right) \dot{\tau} = P  \ , \quad \left(\frac{1}{u^2} + \frac{1}{u_*^2}\right) \dot{x} = L \ , \label{eom1} \\
&& \frac{1}{u^2}\left(\frac{1}{1-\frac{u^2}{u_*^2}}\right) \dot{u}^2 + \left(\frac{1}{u^2}+\frac{1}{u_*^2}\right) \dot{x}^2 -\frac{1}{u^2}\left(1-\frac{u^4}{u_*^4}\right) \dot{\tau}^2 = 0 \ . \label{eom2}
\end{eqnarray}
Eliminating $\lambda$, we get
\begin{eqnarray}
&& \frac{dx}{du} =  \pm \frac{u_*^2}{\sqrt{u^2 + u_*^2}} \frac{1}{\sqrt{u_*^2 \left( \varepsilon^2 - 1 \right) + u^2 }} \ ,  \label{null1} \\
&& \frac{d\tau}{du} =  \pm \frac{\varepsilon u_*^4 \sqrt{u^2 + u_*^2}}{\left( u_*^4 - u^4 \right)} \frac{1}{\sqrt{u_*^2 \left( \varepsilon^2 -1 \right) + u^2}} \ , \quad \varepsilon= \frac{P}{L}  \ , \label{null2}
\end{eqnarray}
with the boundary conditions that $x(0) = x_0$ and $\tau(0) = \tau_0$. The general solution of the above equations is given by
\begin{eqnarray}
\tau(u)_{\pm} & = & \tau_0 \pm  \frac{\varepsilon u_*}{\sqrt{1 - \varepsilon^2}} \ \Pi \left(-1 ; i \sinh^{-1} \left( \frac{u}{u_s}\right) \big| \frac{1}{ \varepsilon^2 - 1}\right) \ , \label{nullsol1} \\
x(u)_{\pm} & = & x_0 \pm  \frac{u_* }{\sqrt{1 - \varepsilon^2}} \ F \left(i \sinh^{-1}\left(\frac{u}{u_s}\right)\big | \frac{1}{\varepsilon^2-1}\right) \ , \label{nullsol2}
\end{eqnarray}
where $F$ and $\Pi$ are Elliptic functions of the first and the third kind, respectively. Here $\{\tau_{\pm}(u), x_{\pm}(u)\}$ generate $\partial_{\pm} \left( \blacklozenge_{\A} \right)$. The corresponding CHI construction will lead to the diagram schematically shown in fig.~\ref{chipic}.

For purely radial null geodesics, which is obtained from the above result by setting $\varepsilon \to \infty$, sent from {\it e.g.}~the future tip of $\lozenge_\A$ we get:
\begin{eqnarray}
\tau_{-} = a - \frac{u_*}{\sqrt{2}} \tanh^{-1} \left( \frac{\sqrt{2} u}{\sqrt{u^2 + u_*^2}} \right)  \ . 
\end{eqnarray}
Setting $\tau_{-} = 0$ yields the minimal radial extent of the bulk causal wedge. This is easily obtained to be:
\begin{eqnarray}
u_{\Sigma} = u_* \tanh\left( \frac{\sqrt{2} a}{u_*} \right) \left[2 - \tanh^2 \left( \frac{\sqrt{2} a}{u_*} \right) \right]^{-1/2}  \ , \label{cwmin}
\end{eqnarray}
which approaches $u_*$ exponentially in the limit $(u_*/a) \to 0$, {\it i.e.}~when $a T_{\rm eff} \gg 1$. In the limit $(u_*/a) \to \infty$, {\it i.e.}~when $a T_{\rm eff} \ll 1$, we get $u_\Sigma \to a$.

We can, given a boundary length-segment of width $(2a)$, now compare the minimal radial extent reached by the bulk causal wedge and a spacelike geodesic associated with the length-segment. For this, we need to compare $u_\Sigma$ in (\ref{cwmin}) with $u_{\rm c}$ in (\ref{uceqn}). It can be checked numerically that for all values of $a$, $u_\Sigma < u_{\rm c}$, {\it i.e.}~in other words, the geodesic probes deeper than the causal wedge. This in accord with general arguments provided in \cite{Hubeny:2012wa}.

In light of \cite{Czech:2012bh, Hubeny:2012wa}, we will now comment on the so called ``entanglement wedge", following the notation and discussions of \cite{Gentle:2015cfp} closely. This has been conjectured to be the {\it most natural} bulk region corresponding to the reduced density matrix of a region in the boundary theory. We will only comment on the AdS$_3$-osm case here, since many of the computations are analytical. Even though we do not know whether the Ryu-Takayanagi prescription for computing entanglement entropy is meaningful in the osm-background, in the same spirit as computing spacelike geodesics that may correspond to the boundary theory correlation function, we will proceed with the computation.

Suppose now, we begin with an interval $\A$ in the boundary. The corresponding co-dimension two extremal area surface (in other words, the spacelike geodesic), denoted by $\E_{\A}$, along with $\A$ defines a co-dimension one surface $\Sigma$, such that $\partial \Sigma = \A \cup \E_{\A}$. The entanglement wedge, subsequently denoted by $\W_{\A}$, is defined as the causal development of $\Sigma$. We will be brief in our discourse, see {\it e.g.}~\cite{Gentle:2015cfp} for more details. To construct the causal development, one can set up the light-sheets given the co-dimension two surface $\E_{\A}$, that are designed using null geodesics emanating from each point on $\E_{\A}$ and orthogonal to $\E_{\A}$ itself.\footnote{From each point on $\E_{\A}$, four null rays emanate: two propagate towards the boundary, one past-directed and the other future-directed, two others propagate towards the bulk. By constraining the expansion $\theta \le 0$, we pick out two sets of null rays that finally comprise the light-sheets associated with $\E_{\A}$.}

Let us now define: $\A = \{(\tau, x) | \tau = 0 , x\in [-a,a]\}$. The co-dimension two surface $\E_{\A}$ is then solved by setting $\tau = 0$ in (\ref{defgeoex}) (for $\kappa=1$), which in turn implies $P=0$ {\it via} equation (\ref{cons}). The solution of the corresponding extremal surface (geodesic) is:
\begin{eqnarray}
&& x_{\pm}(u) = \pm \frac{L u_*^2}{\sqrt{L^2 u_*^2-2}} F \left(i \sinh ^{-1}\left(u_* \sqrt{\frac{L^2 u_*^2-2}{\left(1-L^2 u_*^2\right) u^2 + u_*^2}}\right)| \frac{L^2 u_*^2}{L^2 u_*^2-2}\right) + x_0 \ , \nonumber\\  \label{geosol} 
\end{eqnarray}
where $F$ is Elliptic function of the first kind and $x_0$ is an integration constant that we need not specify. For simplicity, we can work with $x_+(u)$ here onwards. First, observe that $x_{+}(0) - x_{+}(u_{\rm c}) = x_{+}(0) = a$, in which $u_{\rm c}$ is the turning point of the geodesic, and thus relates $L$ with $a$.

We want to find null geodesics that are orthogonal to (\ref{geosol}). Suppose now, the tangent vector to (\ref{geosol}) is denoted by $\Xi^\mu = \{0, \Xi^x, \Xi^u\}$ and the null normal by $\N^\mu= \{\N^\tau, \N^x, \N^u\}$; then
\begin{eqnarray}
\Xi^\mu \N_\mu = 0 \ , \quad \N^\mu \N_\mu = 0 \ .
\end{eqnarray}
We find $\Xi^x / \Xi^u = dx_+ / du$ by using (\ref{geosol}) to be:
\begin{eqnarray}
\frac{\Xi^x}{\Xi^u} = -   \frac{L u u_*^3}{\sqrt{\left(u_*^4 - u^4\right) \left(u^2 \left(1 - L^2 u_*^2\right) + u_*^2\right)}} \ .
\end{eqnarray}
Using the above and solving the orthogonality condition between $\Xi^\mu$ and $\N^\mu$ we get:
\begin{eqnarray}
\frac{\N^x}{\N^u} = - \left. \frac{ \left(u^2 + u_*^2\right) \sqrt{\left( u_*^4 - u^4\right) \left(-L^2 u^2 u_*^2 + u^2 + u_*^2\right)}}{L u^5 u_*^3 \left(u^2 - u_*^2\right)} \right |_{u=u_{\rm s}}\ , \label{nullrel1}
\end{eqnarray}
where $u_{\rm s}$ is the point of intersection of the null geodesic and the spacelike geodesic. Note that, in obtaining the above, we have also chosen the sign that corresponds to null rays propagating towards the boundary. The null-ness condition yields:
\begin{eqnarray}
\frac{\N^\tau}{\N^u} = \pm \left. \P\left( u, u_*, a \right) \right |_{u = u_{\rm s}}\ , \label{nullrel2}
\end{eqnarray}
where $\P$ is known analytically, the explicit form of which is not particularly illuminating. In writing the above, we have traded $L$ with $a$. Also note that the $\pm$ sign above corresponds to null rays propagating towards boundary in the future and in the past, respectively.

Evidently, the null normal $\N$ is tangent to a null geodesic intersecting the spacelike geodesic (\ref{geosol}) at the radial position $u_{\rm s}$. To construct a null ray that passes through this point and continues towards the boundary, we need to solve for null geodesics in (\ref{eom1}) and (\ref{eom2}). Equivalently one can solve the system (\ref{null1}) and (\ref{null2}) and obtain solutions of the form written in (\ref{nullsol1}) and (\ref{nullsol2}), subject to the boundary condition that we clarify below.

First, we can evaluate the constant $\varepsilon = P / L$ for a null geodesic that is tangent to (\ref{geosol}) at the radial location $u_{\rm s}$. This can be achieved by first observing that, upon using the equations of motion, 
\begin{eqnarray}
\frac{\N^x}{\N^\tau} = \left. \frac{\dot{x}}{\dot{\tau}} = \frac{1}{\varepsilon} \left( 1 - \frac{u^2}{u_*^2} \right) \right |_{u = u_{\rm s}} \ . \label{nullrel3}
\end{eqnarray}
Alternatively, using (\ref{nullrel1}) and (\ref{nullrel2}), we can compute $\N^x / \N^\tau$ which is a function of $\{u_{\rm s}, u_* , a\}$. Combining this with (\ref{nullrel3}), we can express $\varepsilon \left(u_{\rm s}, u_*, a \right)$. We do not explicitly write down the expression here, since the functional forms are algebraically complicated and not particularly illuminating. Furthermore, in order for the null geodesic to intersect the spacelike geodesic in (\ref{geosol}), we demand that at the radial location $u_{\rm s}$, and $\tau = 0$, we have $x = x_+(u_{\rm s})$, where $x_+(u_{\rm s})$ is given by (\ref{geosol}). In brief, the light-sheets are described essentially by the solutions in (\ref{null1}) and (\ref{null2}) subject to the boundary conditions discussed above, and they are qualitatively similar to the ones obtained in a BTZ-background.

\section{Stringy embedding of AdS$_3$}

The standard way, in which bulk AdS$_3$ emerges from a stringy construction is to consider the D$1$-D$5$ bound state.\footnote{For review articles, see {\it e.g.}~\cite{Mandal:2000rp, Kraus:2006wn}. Also note that, AdS$_3$-geometry can arise from wrapping M$5$-branes on a $4$-cycle inside a six dimensional manifold, denoted by ${\cM}_6$, which can be $T^6$, $K3 \times T^2$ or Calabi-Yau three fold. See {\it e.g.}~\cite{Maldacena:1997de} for more details.} The relevant part of the supergravity action, in string frame, is given by
\begin{eqnarray}
S_{\rm sugra} = \frac{1}{2 \kappa^2} \int d^{10} x \sqrt{-G} e^{- 2 \phi} \left( R - 4 \left(\partial \phi\right)^2 - \frac{1}{2} \left| G_3 \right|^2\right) \ .
\end{eqnarray}
The solution is given by
\begin{eqnarray}
ds^2 & = & \left(H_1 H_5\right)^{-1/2} \left( - dt^2 + dx^2 \right)  + \left(H_1 H_5\right)^{1/2} ds_{\mathbb R^4}^2 + \left(\frac{H_1} {H_5}\right)^{1/2} ds_{\cM _4}^2 \ , \\ 
H_{1,5} & = & 1 + \frac{Q_{1,5}}{r^2} \ , \quad  Q_1 = \left(2 \pi \right)^4 \frac{g_s N_1}{V_4} \alpha'^3 \ , \quad Q_5 = g_s N_5 \alpha' \ ,   \\ 
G_3 & = & 2 Q_5 \epsilon_3 + 2 i Q_1 e^{-2\phi} \ast_6 \epsilon_3 \ , \\
e^{- 2 \phi} & = & \frac{H_5}{H_1}  \ .
\end{eqnarray}
Here $\epsilon_3$ is the volume form on the $3$-sphere, $\ast_6$ is the Hodge dual in six dimensions and $N_1$ and $N_5$ are related to the number of D$1$ and D$5$-branes. Finally, ${\cM}_4$ typically represents $T^4$ or $K3$-manifold.

Correspondingly, the near-horizon limit yields, in which we also take $g_s \to 0$ with $g_s N_1 \gg 1$ and $g_s N_5 \gg 1$,
\begin{eqnarray}
ds^2 & = & \frac{r^2}{L^2} \left(- dt^2 + dx^2 \right) + \frac{L^2}{r^2} dr^2 + L^2 d\Omega_3^2 + \left(\frac{Q_1}{Q_5}\right)^{1/2} ds_{\cM_4}^2 \ , \\
G_3 & = & 2 Q_5 \left( \epsilon_3 + i \ast_6 \epsilon_3 \right) \ , \\
e^{- 2 \phi} & = & \frac{Q_5}{Q_1} \ , \quad L^2 = \left(Q_1 Q_5\right)^{1/2} \ .
\end{eqnarray}
The D$1$-branes span $\{t, x\}$-directions and the D$5$-branes wrap $\{t, x\}$ and the ${\cM}_4$. The resulting gauge theory is a $(1+1)$-dim CFT with $\cN = (4,4)$ supersymmetry and the gauge group $U(N_1) \times U(N_5)$. There are hypermultiplet fields transforming as the adjoints of $U(N_1)$ and $U(N_5)$ separately and also as the bifundamentals of $U(N_1) \times U(N_5)$.

Now, one simple way to introduce flavours in this picture is to consider an additional probe D$5'$-branes which extend along $\{t, x, r\}$-directions and wrap the entire $3$-sphere, where $'$ denotes that these are different from the background D$5$-branes. Since this probe sector extends along the radial direction, the gauge symmetry on the probe gets promoted to a global symmetry. The corresponding $(1, 5') / (5',1)$ strings have $4$ Neumann-Dirichlet boundary conditions, and $(5,5') / (5',5)$ strings have $8$ Neumann-Dirichlet boundary conditions.\footnote{As discussed in \cite{Mandal:2000rp}, one can be more precise about the details of the flavour degrees of freedom, however, for our purposes these details do not matter and hence we will not discuss this further.} It is now straightforward to argue that the fundamental sector will be described by a Dirac-Born-Infeld action, and there will be no Wess-Zumino term even when we turn on the electric field to induce the steady-state.

A natural question is whether the probe, thus embedded in the background is a stable one. We will now argue that it indeed is. The argument is by a straightforward computation. One can work with a generic embedding profile for the probe D$5$, parametrized by $z(t, r)$, where $z$-denotes a direction along the $\cM_4$, which we can take to be $T^4$ for simplicity. The corresponding DBI-Lagrangian will take the following form
\begin{eqnarray}
\L \sim r \left( 1 + \alpha r^2 z'^2 - \frac{\alpha \dot{z}^2}{r^2} \right)^{1/2} \ ,
\end{eqnarray}
where $\alpha$ is a constant (that depends on $Q_1$ and $Q_5$) which will not be relevant for our discussions. Also, $' \equiv \partial/ \partial r$ and $\dot{} \equiv \partial/ \partial t$. The general equation of motion resulting from this Lagrangian takes the form
\begin{eqnarray}
\frac{d}{dt} \frac{\partial \L}{\partial \dot{z}} + \frac{d}{dr} \frac{\partial \L}{\partial z'} = 0 \ . \label{d5ind1d5}
\end{eqnarray}
It is easy to check that static solutions of (\ref{d5ind1d5}) are given by: $z_{(0)} = {\rm const}$. Suppose now, we linearize around this classical solution using the following ansatz:
\begin{eqnarray}
z = z_{(0)} + \delta z(t, r) \ .
\end{eqnarray}
The resulting linearized equation of motion allows one to perform a separation of variable, such that $\delta z (t, r) = Y(t) X(r)$, that obeys:
\begin{eqnarray}
\ddot{Y} = - k^2 Y \ , \quad  r^3 \left( 3 X' + r X'' \right) +k^2 X = 0 \ .
\end{eqnarray}
The correct solution is given by
\begin{eqnarray}
Y(t) \sim e^{- i k t} \ , \quad X(r) \sim \frac{1}{k r} H_{1}^{(1)} \left( \frac{k}{r} \right) \ ,
\end{eqnarray}
where $H_1^{(1)}$ is the Hankel function of the first kind. It is easy to check that this solution has the desired ingoing boundary condition at the Poincar\'{e} horizon and normalizability at the conformal boundary. Thus, the fluctuations around $z_{(0)}$ do not grow unbounded and the corresponding classical profiles remain stable.

Now, to read off the gauge theory coupling we can carry out the following exercise. There are two sets of branes in the background, so we should be able to define two sets of coupling constants. However, since the dual gauge theory is $(1+1)$-dimensional, it is natural to define the gauge coupling in terms of the D$1$-branes. To that end, imagine introducing a {\it probe} D$1$-brane along $\{t, x\}$-directions, which will have the following action
\begin{eqnarray}
S_{{\rm D}1} = - T_{{\rm D}1} \int e^{-\phi} \sqrt{- G_{tt} G_{xx}} & = & - T_{{\rm D}1} \left(\frac{Q_5}{Q_1}\right)^{1/2} \frac{1}{L^2} \int \ldots \nonumber \\
& = & - \frac{1}{g_{\rm YM}^2 \alpha'^2} \int \ldots \ . 
\end{eqnarray}
This gives
\begin{eqnarray}
g_{\rm YM}^2 = \left( 2 \pi g_s \right) \alpha'^{-1} Q_1 = \frac{\left( 2 \pi \right)^5}{V_4} g_s^2 \alpha'^2 N_1 \ ,
\end{eqnarray}
where we have used the relation 
\begin{eqnarray}
T_{{\rm D} p } = \frac{1}{g_s} \frac{1}{\left(2 \pi\right)^p} \frac{1}{\alpha'^{(p+1)/2}} \ .
\end{eqnarray}
On the other hand, if we used a D$5$-probe instead, we would get
\begin{eqnarray}
g_{\rm YM}^2 = \left( 2 \pi \right)^5 g_s \alpha' Q_5  = \left( 2 \pi \right)^5 g_s^2 \alpha'^2 N_5 \ .
\end{eqnarray}
Correspondingly, up to a numerical constant, we can define two 't Hooft coupling:
\begin{eqnarray}
\lambda_1 & = & g_{\rm YM}^2 N_1 \sim g_s^2 \alpha'^2 N_1^2 \ , \\
\lambda_2 & = & g_{\rm YM}^2 N_5 \sim g_s^2 \alpha'^2 N_5^2 \ .
\end{eqnarray}
There may be more ways to introduce flavours, and we may enlist those. But it is not important for our discussions.

\section{Acknowledgements}

We would like to thank Amit Ghosh and Tom Hartman for conversations on the energy conditions. SK is supported by NSF grant PHY-1316222. We would like to thank the people of India for their generous support in research in basic sciences.

\appendix

\section{Kruskal extension of BTZ: a brief account} \label{sec:KrusBTZ}

To compare the previous case with the more standard BTZ-background, we briefly review the corresponding Kruskal extension. Let us start with the AdS$_3$-BTZ background, written as (suppressing the field theory spatial directions):
\begin{eqnarray}
ds^2 = - (r^2 - r_{\rm h}^2 ) dt^2 + \frac{dr^2}{r^2 - r_{\rm h}^2}  \ .
\end{eqnarray}
Now, the tortoise coordinate is:
\begin{eqnarray}
r_* & = & - \frac{1}{r_{\rm h}} \tanh^{-1} \left( \frac{r}{r_{\rm h}} \right)  \ , \quad \forall r \in [r_{\rm h}, \infty] \ . \label{limexact} \\
r_* & = & - \frac{1}{r} - \frac{r_{\rm h}^2}{3 r^3} + \ldots \ , \quad r \to \infty \ , \label{lim1} \\
r_* & = & \frac{1}{2 r_{\rm h}} \log(r - r_{\rm h}) + \ldots  \ , \quad r \to r_{\rm h} \ . \label{lim2} 
\end{eqnarray}
Now we can define
\begin{eqnarray}
 u = t - r_* \ , \quad v = t + r_* \ , \quad U = - e^{- r_{\rm h} u } = T - X \ , \quad V = e^{r_{\rm h} v} = T + X \ , \label{btzTX}
\end{eqnarray}
which yields
\begin{eqnarray}
ds^2 = - \frac{r^2 - r_{\rm h}^2}{r_{\rm h}^2} e^{- 2 r_{\rm h} r_*} (- dT^2 + dX^2 ) = \Omega(r) (- dT^2 + dX^2 ) \ .
\end{eqnarray}
Using the expansions in (\ref{lim2}), it can be shown that $\Omega(r_{\rm h}) = 2 r_{\rm h}$, which is finite and has an inverse. Thus we can now extend the patch to its original range of $U <0$, $V>0$ to $\{U, V\} \in (-\infty, \infty)$ with the constraint that $0 < r < \infty$.

We can explicitly write down the coordinate transformations as:
\begin{eqnarray}
T^2 - X^2 = e^{2 r_{\rm h} r_*} \ , \quad  \tanh^{-1} \left( \frac{T}{X} \right) = r_{\rm h} \, t \ .
\end{eqnarray}
and also deduce that
\begin{eqnarray}
e^{2 r_{\rm h} r_*} = \frac{r - r_{\rm h}}{r + r_{\rm h} } \ .
\end{eqnarray}
From these definitions it is clear that $r \to 0$ corresponds to $e^{2 r_{\rm h} r_*} =-1 = - (T^2 - X^2)$ and $ r \to \infty$ corresponds to $e^{2 r_{\rm h} r_*} = 1 = - (T^2 - X^2)$.

Let us now determine where the boundary and the singularity intersect the Kruskal coordinates. Along the $r=0$ curve (singularity), setting $X=0$ gives $T_{\rm sing} = \pm 1$. Along the $r=\infty$ curve (boundary) setting $T=0$ gives $X_{\rm bound} = \pm 1$. Since $|T_{\rm sing}| = |X_{\rm bound}|$, the resulting Penrose diagram will be a perfect square,-- as is well-known.

\section{Kruskal extension of osm: A detailed account} \label{sec:osmKrus3}

Here we will provide the details of the coordinate transformations corresponding to the Kruskal extension of osm. Let us take the AdS$_3$-osm, represented by:
\begin{eqnarray}
ds^2 = - \frac{1}{u^2} \left( 1 - \frac{u^4}{u_{\rm E}^4} \right) d\tau^2 + \left(\frac{1}{u^2} + \frac{1}{u_{\rm E}^2} \right) dx^2  + \frac{1}{u^2} \left( 1 - \frac{u^2}{u_{\rm E}^2} \right)^{-1} \ .
\end{eqnarray}
The radius of curvature has been set to unity. Now define:
\begin{eqnarray}
r = \frac{u_{\rm E}^2}{u} \ , \quad r_{\rm E} = u_{\rm E} \ , \quad t \to \frac{t}{u_{\rm E}^2} \ , \quad x \to \frac{x}{u_{\rm E}^2} \ , 
\end{eqnarray}
to get
\begin{eqnarray}
ds^2 = - r^2 \left( 1 - \frac{r_{\rm E}^4}{r^4} \right) d\tau^2 + \frac{dr^2}{r^2} \left( 1 - \frac{r_{\rm E}^2}{r^2} \right)^{-1} + (r^2 + r_{\rm E}^2) dx^2 \ .
\end{eqnarray}
The tortoise coordinate is now defined as
\begin{eqnarray}
r_* = \int \frac{r dr}{(r^2 - r_{\rm E}^2 ) \sqrt {r^2 + r_{\rm E}^2 }} + {\rm const} \ ,
\end{eqnarray}
which yields
\begin{eqnarray}
r_* & = & - \frac{1}{r} - \frac{r_{\rm E}^2}{6 r^3} + \ldots \ , \quad r \to \infty \ , \label{rs1} \\ 
r_* & = & - \frac{1}{2\sqrt{2} r_{\rm E}} \log (r- r_{\rm E}) + \ldots \ , \quad r \to r_{\rm E} \ , \label{rs2} \\ 
r_* & = & - \frac{1}{\sqrt{2} r_{\rm E}} \coth^{-1} \left( \frac{\sqrt{2} r_{\rm E}}{ \sqrt{r^2 + r_{\rm E}^2}} \right) \ , \quad \forall r \in [r_{\rm E}, \infty]  \label{esexact} \ . \\
{\rm Alternatively} \quad r_* & = & - \frac{1}{2\sqrt{2} r_{\rm E}}  \log \left( \frac{\sqrt{r^2 + r_{\rm E}^2} + \sqrt{2} r_{\rm E}}{\sqrt{r^2 + r_{\rm E}^2} - \sqrt{2} r_{\rm E}} \right) \ .
\end{eqnarray}
So the corresponding range is: $- \infty < r_* < 0$. 

Now define the incoming and outgoing null coordinates:
\begin{eqnarray}
\tilde{u} = \tau - r_* \ , \quad \tilde{v} = \tau + r_* \ , \label{defuv}
\end{eqnarray}
which gives
\begin{eqnarray}
ds^2 = - F(r) d\tilde{u} d\tilde{v} \ , \quad F(r) = r^2 \left( 1 - \frac{r_{\rm E}^4}{r^4} \right) \ .
\end{eqnarray}
We further define
\begin{eqnarray}
U = - e^{-(\sqrt{2} r_{\rm E}) \tilde{u}} \ , \quad V = e^{(\sqrt{2} r_{\rm E}) \tilde{v}} \ , \label{defUV}
\end{eqnarray}
which yields
\begin{eqnarray}
ds^2 = - \frac{F}{2 r_{\rm E}^2} e^{- (2 \sqrt{2} r_{\rm E}) r_*} \, dU dV = \Omega(r) \, dU dV \ .
\end{eqnarray}
Using (\ref{rs2}), it can be checked that $\Omega(r_{\rm E}) = 4 r_{\rm E}$ and is thus non-degenerate. This makes the osm event-horizon a perfectly regular point and we can extend beyond this. Note that, for $- \infty < u, v < \infty$ we have $- \infty < U < 0$ and $0 < V < \infty$. Also
\begin{eqnarray}
UV = - \left( \frac{\sqrt{r^2 + r_{\rm E}^2} + \sqrt{2} r_{\rm E}}{\sqrt{r^2 + r_{\rm E}^2} - \sqrt{2} r_{\rm E}} \right) \ .  \label{prodUV}
\end{eqnarray}
Let us now define the Kruskal coordinates as:
\begin{eqnarray}
&& U = T - X \ , \quad V = T + X \ , \\
&& ds^2 = \Omega(T, X) (- dT^2 + dX^2 ) \ ,
\end{eqnarray}
Using (\ref{defuv}) and (\ref{defUV}) we get: 
\begin{eqnarray}
- e^{(2 \sqrt{2} r_{\rm E}) r_*} & = & T^2 - X^2 \ , \\
\sqrt{2} r_{\rm E} \, t & = & \tanh^{-1} \left( \frac{T}{X} \right) \ . 
\end{eqnarray}
It is also straightforward to show, using the definition in (\ref{esexact}), that
\begin{eqnarray}
e^{(2 \sqrt{2} r_{\rm E}) r_*} = \frac{\sqrt{2} r_{\rm E} - \sqrt{r^2 + r_{\rm E}^2}} { \sqrt{2} r_{\rm E} + \sqrt{r^2 + r_{\rm E}^2} } \ . 
\end{eqnarray}
The boundary is located at $ r = \infty$ and this corresponds to
\begin{eqnarray}
e^{(2 \sqrt{2} r_{\rm E}) r_*} = -1  \quad \implies \quad T^2 - X^2 = - 1 \ .
\end{eqnarray}
The singularity is located at $r = 0$, which corresponds to
\begin{eqnarray}
e^{(2 \sqrt{2} r_{\rm E}) r_*} = \frac{\sqrt{2} -1}{\sqrt{2} + 1} \quad \implies  \quad T^2 - X^2 = \frac{\sqrt{2} -1}{\sqrt{2} + 1} \ .
\end{eqnarray}
The singularity intersects the Kruskal coordinates at
\begin{eqnarray}
X = 0 \ , \quad T_{\rm sing} = \pm \left( \frac{\sqrt{2} -1}{\sqrt{2} + 1} \right)^{1/2} \ .
\end{eqnarray}
The boundary intersects the Kruskal coordinates at
\begin{eqnarray}
T = 0 \ , \quad X_{\rm bound} = \pm 1 \ .
\end{eqnarray}
Since $|T_{\rm sing}| < |X_{\rm bound}|$, the singularity comes closer to the centre of the diagram than the boundary; and hence the singularity, in the Penrose diagram, will bend inwards.

To draw the Penrose diagram, let us start from $\{\tilde{u}, \tilde{v}\}$ coordinates in (\ref{defuv}) and define:
\begin{eqnarray}
U' = \tan^{-1} U \ , \quad V' = \tan^{-1} V \ . 
\end{eqnarray}
The corresponding time-like and space-like coordinates are defined as:
\begin{eqnarray}
T' = \frac{1}{2} \left(V' + U' \right) \ , \quad X' = \frac{1}{2} \left( V' - U ' \right) \ . \label{TXp}
\end{eqnarray}
In order to draw the Penrose diagram, let us identify the following special regions of the global space-time: \\

\noindent {\bf Horizon: } This is located at $r = r_{\rm E}$, so equation (\ref{prodUV}) gives $UV = 0$. So, we either have $V > 0$ with $U = 0$ or we have $V = 0$ with $U < 0$. These correspond to the Future Horizon, denoted by ${\cal H}_+$, and the Past Horizon, denoted by ${\cal H}_-$, respectively. It is easy to check that ${\cal H}_+$ is described by $T' = X'$ and ${\cal H}_-$ is described by $T' = - X'$. \\

\noindent {\bf Space-like Infinity:} The conformal boundary is located at $r \to \infty$, which yields $U V = -1$. This is described by
\begin{eqnarray}
X' = \frac{1}{2} \left( V' - U' \right) = \frac{1}{2} \tan^{-1} \left(\frac{V - U}{1 + UV} \right) = \frac{\pi}{4} \ .
\end{eqnarray}
This corresponds to a vertical line in the Penrose diagram, and is denoted by $i_0$. \\

\noindent {\bf Time-like Infinity:} The future time-like infinity is located at $\tau = + \infty$, which gives $\tilde{u}, \tilde{v} = \infty$. This, in turn, yields: $V' = \frac{\pi}{2}$, $U' =0$ and is denoted by $i_+$. Similarly, the past time-like infinity is located at $\tau = - \infty$, which gives: $U' = - \frac{\pi}{2}$, $V' = 0$, and is denoted by $i_-$. \\

\noindent {\bf Singularity:} The singularity is located at $r=0$, which is described by the following equation: 
\begin{eqnarray}
UV = \frac{\sqrt{2} -1 } {\sqrt{2} + 1} \ .
\end{eqnarray}
Given the explicit coordinate transformations, one can easily translate this equation in the $\{T', X'\}$-plane in which one observes that the singularity ``bows in", as compared to the BTZ-black hole.

\section{Kruskal extension of higher dimensional osm} \label{sec:osmKrus4}

The AdS$_4$-osm line element is 
\begin{equation}
ds^2 = - \frac{1}{u^2}\bigg(1- \frac{u^4}{u_{\rm E}^4}\bigg)d\tau^2 +  \frac{1}{u^2}\bigg(1- \frac{u^4}{u_{\rm E}^4}\bigg)^{-1} du^2 + \frac{1}{u^2}(dx^2+dy^2) \ .
\end{equation}
With $u= \frac{1}{r}$ , $r_{\rm E}=\frac{1}{u_{\rm E}}$ the line element becomes 
\begin{equation}
ds^2= - r^2\bigg(1- \frac{r_{\rm E}^4}{r^4}\bigg)d\tau^2 +  \frac{1}{r^2}\bigg(1- \frac{r_{\rm E}^4}{r^4}\bigg)^{-1}dr^2 + r^2(dx^2 + dy^2) \ .
\end{equation}
We obtain the tortoise coordinate as before:  
\begin{eqnarray}
d\tau & = & \pm \left(1-\frac{r_{\rm E}^4}{r^4}\right)^{-1} \frac{dr}{r^2} = \pm d r_* \ , \\
\implies \quad r_* & = & \frac{1}{4r_{\rm E} } \left[ 2\tan^{-1} \left(\frac{r}{r_{\rm E}}\right) + \log \left(\frac{r-r_{\rm E}}{r+r_{\rm E}} \right) \right]
\end{eqnarray}
So for $r_{\rm E} < r < \infty$, $-\infty< r_*< \frac{\pi}{4 r_{\rm E}}$. Now we define the null coordinates: $\tilde{u} = \tau - r_* $, $\tilde{v} = \tau + r_*$. The range of the null coordinates are: $\tilde{u}, \tilde{v} \in [-\infty, \infty]$. Subsequently, we exponentiate the null coordinates to define: 
\begin{eqnarray}
U = - e^{- 2 r_{\rm E} \tilde{u}} \quad {\rm and} \quad  V = e^{2 r_{\rm E} \tilde{v}} \ . 
\end{eqnarray}
This yields: $U \in [-\infty, 0]$ and $V \in [0, \infty]$. From the definition, we readily get
\begin{equation}
U V  = - \bigg(\frac{r-r_{\rm E}}{r+ r_{\rm E}}\bigg) \exp\left[2\tan^{-1}\left(\frac{r}{r_{\rm E}}\right) \right] \ . \label{UVads4}
\end{equation}
The compact null coordinates can be defined as: $U' = \tan^{-1} \tilde{u}$, $V' = \tan^{-1} \tilde{v}$; and subsequently the space-like and the time-like ones as: $X' = (1/2) \left( V' - U' \right)$, $T' = (1/2) \left(V' + U' \right)$. \\

\noindent {\bf Horizon:}  The horizon is located at $r=r_{\rm E}$. From (\ref{UVads4}) we get: $U V = 0$. Thus, we have either $V > 0$, $U = 0$ or $V = 0$, $ U< 0$. These correspond to the future horizon $\cH_{+}$ and the past horizon $\cH_-$, respectively. As before, $\cH_{+}$ is described by $T' = X'$ and $\cH_-$ is described by $T' = - X'$. \\

\noindent {\bf Singularity:} The singularity is at $r=0$, equivalently, $U V= 1$. This also yields:
\begin{eqnarray}
T' = \dfrac{1}{2}(U' + V') = \dfrac{1}{2}\tan^{-1} \left (\dfrac{U + V }{1 - U V} \right) = \dfrac{\pi}{4} \ . 
\end{eqnarray}
Thus, in the above coordinate system, the singularity is parallel to the space-like direction $X'$. \\

\noindent {\bf Timelike infinity($i_{\pm}$):} The time-like infinities are located at $\tau = \pm \infty$. The positive sign corresponds to $\tilde{u} = \infty = \tilde{v}$ that in turn yields: $V' = \pi/2$, $U' =0$. The negative sign corresponds to $\tilde{u} = - \infty = \tilde{v}$ that in turn yields: $V' = 0$, $U' = - \pi/2$. \\

\noindent {\bf Boundary:} Finally, the boundary is located at $r=\infty$, which translates to $U V =e^{\pi} $. Since the RHS is larger than unity, the boundary will ``bow out" in this case. This is an alternative manifestation of the singularity ``bowing in" for the AdS$_3$-background. To observe the connection directly, let us simply define:
\begin{eqnarray}
&& U '' = e^{-\frac{\pi}{2}} U \ , \quad V'' = e^{-\frac{\pi}{2}} V \\
&& \implies \quad \left. U'' V'' \right|_{\rm boundary} = 1  \ , \quad \left. U'' V'' \right|_{\rm singularity} = e^{- \pi} \ . 
\end{eqnarray}
Thus, expressed in terms of $\{U'', V''\}$, the boundary remains a straight vertical line, while the singularity bows in.

\section{Geodesic Equations} \label{sec:geod}

In this appendix we set up the equations for analyzing geodesics in a given background. First, for generality, consider the following background:
\begin{eqnarray}
ds^2 = G_{\tau\tau} d\tau^2 + G_{uu} du^2 + G_{xx} dx^2 + G_{yy} d\vec{y}_m^2 \ , \label{genosm}
\end{eqnarray}
where $\{G_{\tau\tau}, G_{uu}, G_{xx}, G_{yy}\}$ are all functions of the radial coordinate $u$. The dual $(m+2)$-dimensional field theory lives along the $\{\tau, x, \vec{y}\}$-directions. We have intentionally introduced an anisotropy between the $x$-direction and the $\vec{y}$-directions. The open string metric that we have considered throughout the text is of the general form written in (\ref{genosm}).

To write down the geodesic equations of motion, parametrized by $X^\mu(\lambda)$, we will explicitly use the symmetries in the background. First, from definition:
\begin{eqnarray}
G_{\mu\nu} \frac{dX^\mu}{d\lambda} \frac{dX^\nu}{d \lambda} = \kappa \ , \label{defgeo}
\end{eqnarray}
where $\lambda$ is an affine parameter and $\kappa = -1, 0, 1$ corresponding to timelike, null and spacelike geodesics. The Killing fields, expressed in terms of the Killing vectors as: $\xi = \xi^\mu \partial_\mu$, corresponding to (\ref{genosm}) are given by
\begin{eqnarray}
\xi_{(1)} = \frac{\partial}{\partial\tau} \ , \quad \xi_{(2)} = \frac{\partial}{\partial x} \ , \quad \xi_{(3)_i} = \frac{\partial}{\partial y_i} \ , \quad i = 1, \ldots, m \ .
\end{eqnarray}
Corresponding to each Killing vector, the first integral of motion is given by
\begin{eqnarray}
G_{\mu\nu} \xi^\mu \frac{dX^\nu}{d\lambda} = {\rm const} \ .
\end{eqnarray}
Explicitly written, the resulting conservation laws are:
\begin{eqnarray}
G_{\tau\tau} \frac{d\tau}{d\lambda} = - P \ , \quad G_{xx} \frac{dx}{d\lambda} = L_x \ , \quad G_{yy} \frac{dy_i}{d\lambda} = L_{y_i} \ . \label{cons}
\end{eqnarray}
The equation of motion for the geodesic resulting from (\ref{defgeo}) is:
\begin{eqnarray}
G_{\tau\tau} \left(\frac{d\tau}{d\lambda} \right)^2 + G_{xx} \left(\frac{dx}{d\lambda} \right)^2 + G_{yy}  \left(\frac{dy_i}{d\lambda} \right)^2 + G_{uu}  \left(\frac{du}{d\lambda} \right)^2 =  \kappa \ . \label{defgeoex}
\end{eqnarray}
Using (\ref{cons}), we can rewrite the above equation as:
\begin{eqnarray}
\left(\frac{du}{d\lambda} \right)^2 + V_{\rm eff} = 0 \ , \quad V_{\rm eff} = \frac{1}{G_{uu}} \left[ -\kappa - \frac{P^2}{|G_{\tau\tau}|} + \frac{L_x^2}{G_{xx}} + \frac{L_{y_i}^2}{G_{yy}} \right] \ . \label{geoveff}
\end{eqnarray}
We have used (\ref{geoveff}) in the main text.

\section{Expansion along Null Congruence}

We will discuss only the asymptotically AdS$_3$-case here, closely following \cite{Hubeny:2007xt}. For completeness, the metric is:
\begin{eqnarray}
ds^2 = - \frac{1}{u^2} \left(1 - \frac{u^4}{u_*^4} \right) d\tau^2  + \left( \frac{1}{u^2} + \frac{1}{u_*^2} \right) dx^2 + \frac{1}{u^2} \left( 1 - \frac{u^2}{u_*^2} \right)^{-1} du^2 \ . \label{osm3now}
\end{eqnarray}
Given a co-dimension two surface, described by
\begin{eqnarray}
\varphi_1 = \tau - \tau_0 = 0 \ , \quad \varphi_2 = x - F(u) = 0 \ , \label{codim2}
\end{eqnarray}
which captures a typical profile of the extremal spacelike geodesic, one can construct the corresponding light-sheet by considering null rays emanating from this geodesic. In principle, one can also consider a case in which $\tau_0$ is not a constant, but a function; furthermore the data $\{\varphi_1, \varphi_2\}$ may not even solve any equation of motion. However, here we will merely comment on the case where (\ref{codim2}) represents an extremal geodesic on a constant time-slice.

Generally though, given $\varphi_i$, $i=1,2$ one can immediately construct two null vectors that are orthogonal to the co-dimension two surface:
\begin{eqnarray}
N_{\pm}^\mu \sim \S^{\mu\nu} \left( \nabla_\nu \varphi_1 +  \B_{\pm} \nabla_\nu \varphi_2 \right) \ ,
\end{eqnarray}
where $\B_{\pm}$ are hitherto undetermined. One now imposes the following constraints:
\begin{eqnarray}
N_{-}^\mu N_{- \mu} = 0 = N_{+}^\mu N_{+ \mu} \ , \quad N_{-}^\mu N_{+ \mu} = -1 \ .
\end{eqnarray}
The constraints above determine $\B_{\pm}$ and the overall normalization for the null vectors uniquely.

\begin{figure}[ht!]
\centering
\includegraphics[width=0.8\textwidth]{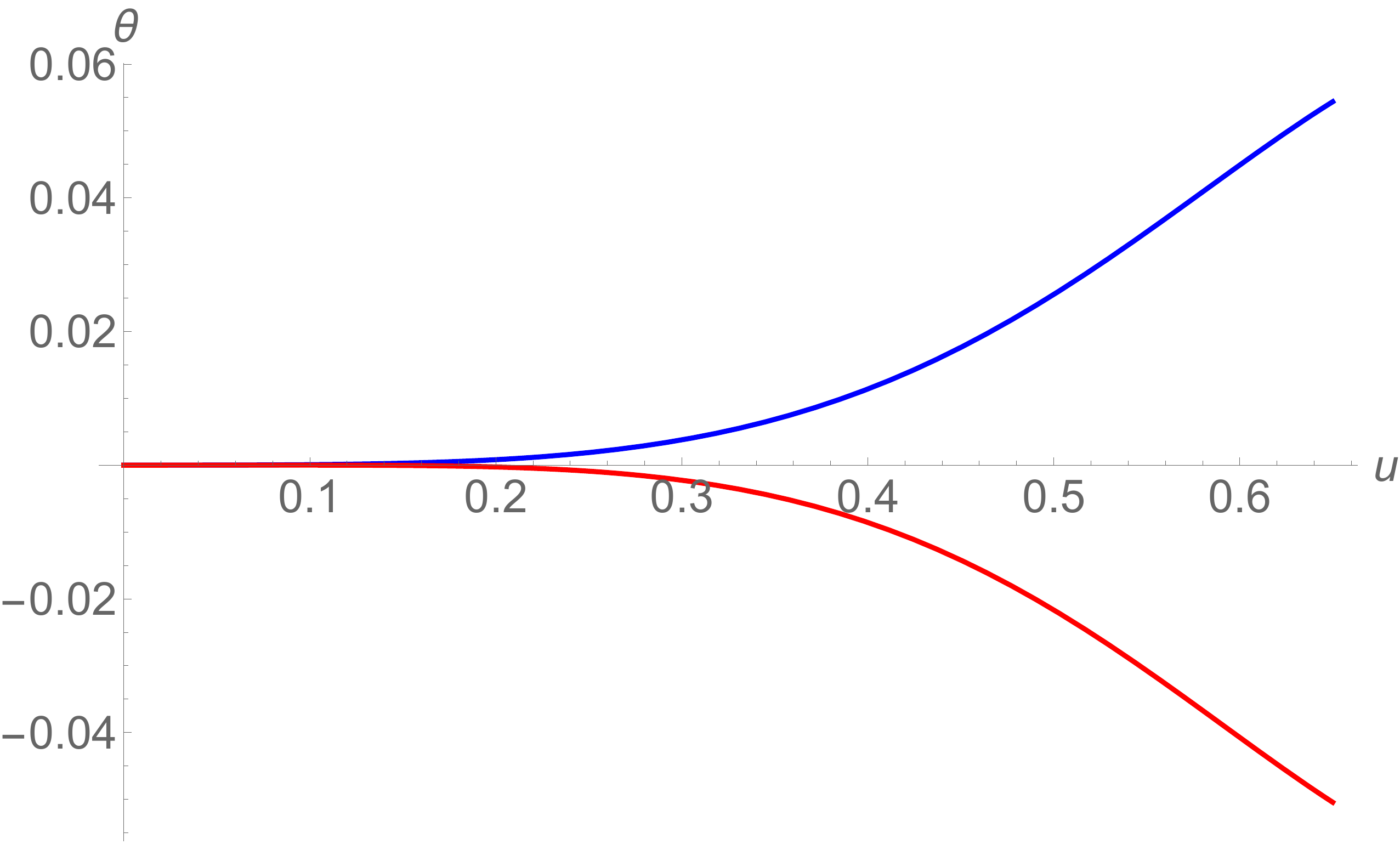}
\caption{The expansion from the co-dimension two surface along the two null normal directions. The blue and red curve corresponds to $\theta_+$ and $\theta_-$, respectively. We have set $u_*=1$ and further $L=3.3$. Corresponding to these choices, there exists an $u_{\rm c}$ up to which $\theta_{\pm}$ has been plotted. }
\label{expanpm}
\end{figure}

In the background (\ref{osm3now}), for an extremal geodesic, the two null normals are:
\begin{eqnarray}
&& N_{\pm} = \left\{ \frac{u u_*^4}{\sqrt{2} \left( u^4 - u_*^4\right) }, \pm \frac{u u_*^2}{\Gamma_1 \left( u^2 + u_*^2 \right) }, \pm \frac{L u \left( u^2 - u_*^2 \right) }{\Gamma_2 u_*^2} \right\} \ , \\
&& \Gamma_1 = \sqrt{2 + \frac{2 L^4}{\left( 1 - \frac{u^4}{u_*^4}\right) \left( \frac{1}{u^2} + \frac{1}{u_*^2} - L^2\right)}} \ , \\
&& \Gamma_2 = \sqrt{2 \left( 1 - \frac{u^4}{u_*^4}\right) \left( \frac{1}{u^2} + \frac{1}{u_*^2} - L^2\right) + 2 L^2} \ .
\end{eqnarray}
To compute the expansion, denoted by $\theta_{\pm}$, from the co-dimension two surface propagated along the null vectors we define the induced metric as:
\begin{eqnarray}
h^{\mu\nu} = \S^{\mu\nu} + N_{+}^\mu N_{-}^\nu + N_{-}^\mu N_{+}^\nu \ ,
\end{eqnarray}
and subsequently compute:
\begin{eqnarray}
\theta_{\pm} = h^{\mu\nu} \nabla_\mu N_{\pm \nu} \ . \label{expantheta}
\end{eqnarray}
A representative plot is shown in fig.~\ref{expanpm}. It is easy to check that the behaviour in fig.~\ref{expanpm} is identical to the one in BTZ-background. 

\end{document}